\documentclass[]{wp-hdsr}
\usepackage{lineno}
\usepackage[inkscapeformat=png]{svg}
\usepackage{algorithm}
\usepackage{algpseudocode}
\usepackage{xspace}
\usepackage{longtable}
\usepackage{color, colortbl}
\usepackage{amsmath}
\usepackage{hyperref}
\hypersetup{colorlinks=true,linkcolor=blue}
\definecolor{Gray}{gray}{0.9}
\definecolor{Medgray}{gray}{0.8}

\graphicspath{{/}}

\begin{document}

\newcommand{\rHDFBT}{$\text{rHDF}_{b,t}$\xspace}
\newcommand{\rHDFB}{$\text{rHDF}_{b}$\xspace}
\newcommand{\chiUS}{161,109,592,812\xspace}
\newcommand{\censusBlocks}{11,078,297\xspace}    
\newcommand{\USpopulation}{308,745,538\xspace}
\newcommand{\blocksUS}{6,207,027\xspace}  
\newcommand{\censusTracts}{73,057\xspace} 
\newcommand{\tractsUS}{72,531\xspace} 
\newcommand{\censusCounties}{3,143\xspace}

\newgeometry{bottom=1.5in}


\volumeheader{Just Accepted}{}{10.1162/99608f92.4a1ebf70}

 
\begin{center}
  \title{A Simulated Reconstruction and Reidentification Attack on the 2010 U.S. Census}
  \maketitle

  \thispagestyle{empty}
  

  \begin{tabular}{cc}
    John M. Abowd\upstairs{\affilone,\affilseven}, 
    Tamara Adams\upstairs{\affilseven}, 
    Robert Ashmead\upstairs{\affilthree,\affilnine},
    David Darais\upstairs{\affilfour}, \\
    Sourya Dey\upstairs{\affilfour}, 
    Simson L. Garfinkel\upstairs{\affilfive,\affilnine},
    Nathan Goldschlag\upstairs{\affileight,\affilnine}, 
    Michael B. Hawes\upstairs{\affiltwo,*}\\
    Daniel Kifer\upstairs{\affiltwo}, 
    Philip Leclerc\upstairs{\affilnine},
    Ethan Lew\upstairs{\affilfour}, 
    Scott Moore\upstairs{\affilfour}, \\
    Rolando A. Rodríguez\upstairs{\affiltwo}, 
    Ramy N. Tadros\upstairs{\affilfour},
    Lars Vilhuber\upstairs{\affilone,\affilnine}\\
   {\small \upstairs{\affilone} Cornell University} 
   {\small \upstairs{\affiltwo} U.S. Census Bureau} 
   {\small \upstairs{\affilthree} Ohio Colleges of Medicine Government Resource Center} \\
   {\small \upstairs{\affilfour} Galois, Inc.} 
   {\small \upstairs{\affilfive} BasisTech and Harvard} 
   {\small \upstairs{\affilseven} U.S. Census Bureau (retired)} \\
   {\small \upstairs{\affileight} Economic Innovation Group} 
   {\small \upstairs{\affilnine} Formerly U.S. Census Bureau} \\

  \end{tabular}
  
  \emails{
    \upstairs{*}Corresponding author: michael.b.hawes@census.gov 
    }

\begin{abstract}

For the last half-century, it has been a common and accepted practice for statistical agencies, including the United States Census Bureau, to adopt different strategies to protect the confidentiality of aggregate tabular data products from those used to protect the individual records contained in publicly released microdata products. This strategy was premised on the assumption that the aggregation used to generate tabular data products made the resulting statistics inherently less disclosive than the microdata from which they were tabulated. Consistent with this common assumption, the 2010 Census of Population and Housing in the U.S. used different disclosure limitation rules for its tabular and microdata publications. This paper demonstrates that, in the context of disclosure limitation for the 2010 Census, the assumption that tabular data are inherently less disclosive than their underlying microdata is fundamentally flawed. The 2010 Census published more than 150 billion aggregate statistics in 180 table sets. Most of these tables were published at the most detailed geographic level---individual census blocks, which can have populations as small as one person. Using only 34 of the published table sets, we reconstructed microdata records including five variables (census block, sex, age, race, and ethnicity) from the confidential 2010 Census person records. Using only published data, an attacker using our methods can verify that all records in 70\% of all census blocks (97 million people) are perfectly reconstructed. We further confirm, through reidentification studies, that an attacker can, within census blocks with perfect reconstruction accuracy, correctly infer the actual census response on race and ethnicity for 3.4 million vulnerable population uniques (persons with race and ethnicity different from the modal person on the census block) with 95\% accuracy. Having shown the vulnerabilities inherent to the disclosure limitation methods used for the 2010 Census, we proceed to demonstrate that the more robust disclosure limitation framework used for the 2020 Census publications defends against attacks that are based on reconstruction. Finally, we show that available alternatives to the 2020 Census Disclosure Avoidance System would either fail to protect confidentiality, or would overly degrade the statistics' utility for the primary statutory use case: redrawing the boundaries of all of the nation's legislative and voting districts in compliance with the 1965 Voting Rights Act. This is the accepted \emph{Harvard Data Science Review paper}. The accepted supplemental text is here: \url{https://arxiv.org/abs/2312.11283v2}.
\end{abstract}
\end{center}


\copyrightnotice

\section*{Media Summary}
The 2010 Census tabular data publications were protected by a combination of swapping geographic identifiers and aggregating statistics to the census block. The 2010 Census microdata publications were protected by sampling and restricting geographic identifiers to areas with populations of at least 100,000. The 2020 Census used a differential privacy framework to produce both tabulations and microdata from the same set of protected statistics. The methods used to prevent disclosure of confidential information in the 2010 Census failed because the tabular statistics could be accurately converted back to microdata via reconstruction. The reconstructed microdata failed the 2010 Census disclosure limitation rules: they were not a sample, and they contained the geographic identifier for a census block, which can have a population as small as one person. The use of weaker disclosure limitation standards for tabulations as compared to microdata is precisely the flaw that our attack exploits and the recommendation that our research challenges. Thus, the 2010 Census disclosure limitation rules could not be used for the 2020 Census, necessitating rapid research and development of a new disclosure avoidance system for balancing privacy and utility. This paper also shows how the potential choices for the 2020 Census disclosure avoidance system---suppression, enhanced swapping, and differential privacy---addressed the risks exposed by our reconstruction and reidentification studies. Among the available alternatives, only the 2020 Census Disclosure Avoidance System successfully defeats the reconstruction-based attacks while preserving the accuracy required for redistricting data.

\restoregeometry
\newgeometry{bottom=0.5in}
\section{Introduction}

Statistical agencies, like the United States Census Bureau, use a variety of statistical disclosure limitation (SDL) methods to protect the confidentiality of data subjects' information when producing and publishing statistical products. These confidentiality protections are both legally mandated \citep{title13} and necessary to maintain the public's trust that their information will only be used for statistical purposes \citep{cipsea}, as now embodied in the trust regulation \citep{OMB:2024}. 

In the context of recurring statistical product releases, like the Decennial Census of Housing and Population (hereinafter, ``the decennial census'' or ``the 20XX Census'' in reference to a particular decade's decennial census), responsible application of SDL requires evaluation of each technique's effectiveness as it is being designed, but also evaluation of the ongoing effectiveness of the techniques over time as the disclosure risk landscape changes. These periodic assessments of the residual disclosure risk of already implemented SDL frameworks serve a function very similar to red hat penetration tests commonly used to assess the ongoing effectiveness of organizations' cybersecurity infrastructure. The objectives of these evaluations are straightforward: Do the protections being employed achieve the overall objectives they were intended to, or have those protections eroded in the face of new or evolving types of attack?

One such periodic re-evaluation of SDL methods by the Census Bureau motivated the results included in this paper. Specifically, as this paper will show, the suite of SDL methods used by the Census Bureau to protect the 2010 Census failed to achieve the overall degree of protection those methods were intended to provide.   

\subsection{SDL methods and standards adopted for the 2010 Census}
\label{subs:SDL_2010}

Data from the 2010 Census were published and made available to the public in two primary forms: tabulations (counts of persons or households that share one or more characteristics, like ``Number of people residing in California who are White alone, male, and 26 years old'') and microdata (individual-level census responses released as part of a ``public-use microdata sample'').\footnote{In Census Bureau nomenclature, White alone means the individual responded White on the race question and did not select any of the other five choices. Similarly, Asian alone means the individual selected only the Asian race category. Finally, Hispanic or Latino means that the individual selected that answer on the separate ethnicity question. In this paper, we will use the shorthand ``Hispanic'' for such responses.} As was common practice across the statistical community \citep{abowd:hawes:2024}, the Census Bureau adopted less stringent SDL requirements for tabulations than for microdata products.  

The primary SDL framework underlying all published 2010 Census products (both tabular and microdata) was household-level record swapping \citep{mckenna:2018:disclosure,mckenna:2019:microdata} applied to the 2010 Census Edited File (CEF)---the collected, edited, and imputed census responses that are considered the archival record of this decennial census. The resulting swapped file, known as the Hundred Percent Detail (HDF) file, contained \USpopulation\footnote{Throughout this paper, when statistics provided within the text are based on publicly available sources, those statistics are published at full precision. When statistics are derived from confidential sources, they are rounded to four significant digits, consistent with applicable Census Bureau Disclosure Review Board rounding rules.} microdata records (one for every person enumerated in the 2010 Census).  

Aggregate tabulations from the HDF were approved for release at geographic levels as fine as the individual census block (the most granular level of geographic detail, a primitive geographic unit with precisely defined geographical boundaries) without additional SDL protections \citep{mckenna:2018:disclosure}. In 2010, the U.S. was subdivided into \censusBlocks mutually exclusive census blocks that tessellated the country; of these \blocksUS had positive populations ranging from one person to more than 1,000 persons. 

Acknowledging the presumed higher disclosure risk associated with the microdata records in the HDF, the Census Bureau designated the HDF itself to be confidential and required the application of additional SDL protections for any microdata products derived from the HDF. These additional protections for microdata products included restricting those products to a 10 percent sample of the HDF, synthesizing values on some records, top-coding selected variables, perturbing age for large households, reducing detail on some categorical variables, and most notably, limiting geographic precision of the microdata products to areas with a minimum population of 100,000 persons \citep{mckenna:2019:microdata}. See also \citet[p. 7-6]{sf1:2010:tech} and \citet[2-1, 2-2]{PUMS:2010:tech}. 

\subsection{Motivation and organization of this paper}

The differing requirements for tabular and microdata products derived from the confidential 2010 Census HDF are inconsistent if the published tabular data can be used to reconstruct an accurate image of the underlying confidential microdata. If an accurate reconstruction of record-level HDF microdata is possible from tabular summaries, then the SDL rules adopted for the 2010 Census should have required that the same stringent disclosure limitation procedures used for the published microdata products also be applied to the published tabular summaries. 

Furthermore, the requirement that more stringent SDL methods be used for microdata products derived from the HDF reflects the Census Bureau's understanding \textit{prior to the 2010 Census} of the inherent, residual disclosure risk of the swapped microdata records contained in the 2010 Census HDF. Successful linkage of these HDF microdata records to an external source of person-level or household-level data containing personal identifiers would result in the potential reidentification of a respondent or of another person in the respondent's household, thus revealing response data associated with that person. 

This method of reidentification is a traditional attack vector that has been recognized by statistical agencies \citep{spwp22,mckenna:2019:reid}, the National Institute for Standards and Technology \citep{garfinkel:near:etal:2023}, and general researchers \citep{rocher:et:al:2019, dick:etal:2023}. Consequently, if an accurate reconstruction of the record-level HDF microdata is possible from the published tabular summaries, it makes the published 2010 Census data susceptible to a reconstruction-abetted reidentification attack. 

With these concerns about the ongoing effectiveness of the 2010 Census SDL protections in mind, this paper sets out to answer the following research questions: (1) \textit{Can individual, confidential microdata records contained in the HDF be accurately reconstructed from the tabular summaries?} (That is, was the Census Bureau wrong to adopt different SDL standards for tabular summaries than for the microdata records from which those summaries were calculated?) and (2) \textit{Can a hypothetical attacker confidently and accurately reidentify specific individuals in microdata records reconstructed from the 2010 Census tables?} (That is, from the opposite perspective, did the record swapping mechanism used to create the HDF adequately protect the HDF's microdata records against disclosure via a traditional reidentification attack?) 

Section \ref{sec:aggregation_hdf} demonstrates that aggregation into tabular summaries is not sufficient to protect the HDF microdata records. It is, in fact, possible to produce an accurate reconstruction of those records from the published 2010 Census tables. In Section \ref{sec:disclosure_risk_hdf}, we proceed to demonstrate that those reconstructed HDF microdata records can be successfully reidentified, revealing confidential response information about those individuals. This shows that the Census Bureau was correct in its determination prior to the 2010 Census that record swapping alone was insufficient to protect the information contained in the HDF. 

Given the Census Bureau's ongoing responsibility to protect confidentiality, the answers to these two research questions necessarily prompt a final question: (3) \textit{Do the new SDL methods that the Census Bureau adopted for the 2020 Census, based on the framework of differential privacy, provide better protection against these types of attack, and if so, do they perform better than feasible alternatives (e.g., enhanced versions of the 2010 record swapping algorithm at varying strengths)?} Section \ref{sec:2020_das} demonstrates that the 2020 Disclosure Avoidance System does, in fact, provide stronger protections than the techniques used for the 2010 Census, and that it does so more effectively and with less degradation of data quality than the feasible alternatives.

In answering these questions, we provide the first demonstration by a national statistical agency that the reconstruction predicted by \citet{Dinur:Nissim:2003:RIW:773153.773173} is feasible at scale. We establish that one does not need access to any confidential data (1) to reconstruct a close approximation of the confidential individual HDF microdata records from which the publicly released 2010 Census tables were generated, (2) to know that many of the reconstructed microdata records are identical to their corresponding records in the confidential microdata (using a 38-age-bin schema), and, in many cases, (3) to know exactly which records those are (see Section \ref{sec:solvar_results}). While we acknowledge this confidence in the exact reconstruction of HDF records is strictly true for ``only'' 97 million persons, adding more tables and more features can only increase this number. 

This paper also provides the first empirical demonstration that reconstructed microdata records can be used to successfully and accurately reidentify vulnerable data subjects, a finding that further highlights the disclosure risk vulnerabilities of record swapping-based SDL mechanisms, such as those used to produce the 2010 Census HDF.  In fact, our reconstruction of the swapped HDF microdata records can be used to make high-precision confidentiality-violating inferences about vulnerable data subjects. As shown in Section \ref{sec:disclosure_risk_hdf} the reidentification precision rate on these inferences for nonmodal vulnerable populations is at least 60.3\% and at worst 95.4\%, far in excess of statistical baselines. 

Our contributions are important because traditional disclosure limitation experts continue to dispute the efficacy of reconstruction-based attacks \citep{Muralidhar:2022,muralidhar:domingo-ferrer:2023} using incomplete formulations of the problem, and domain experts continue to assert that the methods are no better than guessing \citep{ruggles:vanriper:2022,francis:2022} or ineffective \citep{kenny:et:al:2021,ruggles2024privacy}. Many of these critiques are addressed directly in \citet{jarmin:etal:2023} and \citet{garfinkel:2023}. 

The analysis of how to properly assess the disclosure risk associated with publishing massive quantities of tabulations from a single confidential input continues to focus on methods with the same flaws that our experimental attack exploits \citep{hotzetal:2022}. Major textbooks and review articles on SDL \citep{willenborg:deWaal:2000,hundepool:et:al:2012,elliot:domingo-ferrer:2018,duncan:etal:2011:statistical} continue to recommend using distinct methods for tabular and microdata publications. However, in the context of large-scale statistical products, the format of the data publication is immaterial because, as we show in this paper, tabulations can be converted into microdata, and an attacker can, using only publicly available information, derive high-quality bounds on how closely the microdata created in this way resemble the microdata used to calculate the tabulations. The use of weaker SDL standards for tabulations as compared to microdata is precisely the flaw that our attack exploits and the recommendation that our research challenges.  

Throughout this paper we cross reference our supplemental text, available at\\  https://doi.org/10.48550/arXiv.2312.11283, which provides detailed commentary, mathematics, and additional statistical results that further support our analyses. 

\section{Can individual, confidential microdata records contained in the HDF be accurately reconstructed from the 2010 Census tabular summaries?}
\label{sec:aggregation_hdf}

The Census Bureau considered the 2010 HDF microdata to be confidential, requiring the application of more stringent SDL to any microdata products derived from the HDF, but allowing tabular summaries of the HDF to be published with geographic precision down to the census block level with no additional SDL beyond aggregation. To evaluate whether this aggregation afforded sufficient protection, we attempted to use a subset of the published 2010 tables to reconstruct the underlying HDF microdata records used to produce those tables. To be clear, a successful reconstruction of the HDF would recreate individual microdata records with the 2010 Census record swapping protections already applied (we will demonstrate in Section \ref{sec:disclosure_risk_hdf} that swapping, as applied in 2010, does not adequately protect against disclosure). The objective of this section is to demonstrate that accurate reconstruction of the swapped HDF microdata records is possible from the tabular summaries.

\subsection{High-level reconstruction methodology}

\emph{Database reconstruction} is a procedure commonly studied in the privacy and confidentiality literature. In this work, we use database reconstruction to mean an attempt to re-create the record-level image of the database from which a set of published query results or tabulations were originally calculated; in this case, that is the confidential HDF.\footnote{See \citet{garfinkel2018a} for a longer discussion of how to understand database reconstruction.} Database reconstruction attempts to reverse-engineer the  confidential HDF records that were the input data used in a tabulation system with the goal of making the reconstruction as close as possible to these confidential data. 

We note that the reconstruction described here is not the most powerful feasible reconstruction because we used only a subset of the 2010 Census tables. In addition, we made no attempt to reconstruct households or household-level information, and we did not use statistical modeling to improve the reconstruction or undo the effects of swapping (e.g., in census blocks where multiple different reconstructions were consistent with the published tables, we did not use statistical methods to identify the most likely reconstruction). Even without these enhancements, in Section \ref{sec:solvar_results} and Table \ref{tab:sol_var_quantiles}, we show that the tabular summaries published for the 2010 Census are highly susceptible to reconstruction.

\subsection{Inputs used for the reconstruction}
\label{sec:inputs_recon}

We describe here only those datasets used in our reconstruction research. The additional data sets used in our reidentification research are discussed in Section \ref{sec:reid_sources}. 

The primary data releases from the 2010 Census are the Redistricting Data (Public Law 94-171) Summary File, which is produced pursuant to 13 U.S. Code § 141(c) to support the re-drawing of geographic boundaries for political districts in each U.S. State, and Summary File 1 (SF1), the main release of demographic data. The redistricting data contain four person-level tables about race and ethnicity cross-tabulated by whether the person is age 18+ (voting age) or younger (not voting age). SF1 contains all tables released in the redistricting data without any changes; however, those tables are renamed and renumbered in SF1. We refer to all tables in this paper by their SF1 names and titles. 

Table \ref{tab:SF1_tables} shows the 34 tables from the 2010 Census SF1 that we included in our reconstruction. These 34 tables were computed from records about persons in the HDF, called the ``population universe'' in technical documentation. We specifically did not use tables where the universe was households, which means that we did not use the ``relationship to the householder'' information to reconstruct household characteristics or to improve the reconstructed data for persons in those households.

These 34 SF1 tables contain multidimensional marginal counts related to sex, age, race, and ethnicity by census block and census tract. Our reconstructed microdata thus contain these variables and only these variables, which we call the feature sets  \{\textit{block, sex, age, race, ethnicity}\} for reconstructions based on both tract- and block-level tables and  \{\textit{block, sex, agebin, race, ethnicity}\} for reconstructions based on block-level tables only.\footnote{Throughout this article and the supplemental text, we italicize block, sex, age, race, ethnicity, and other features of our analysis datasets to indicate the precise coding schema used in the 2010 Census. This avoids confusion with the ordinary English meaning of the words. See Table 1 in the supplemental text for the inventory of these features.} The reconstructed data, therefore, necessarily reflect the schemas used for SF1 and are only informative about variables, in particular age, in those schemas used for publication. (See the supplemental text, Section 3.4 for additional details.)

\begin{table}[ht]
    \caption{Tables from 2010 Summary File 1 Used in Reconstruction Experiments}    
    \label{tab:SF1_tables}
    \centering
    \begin{tabular}{l l } 
        \multicolumn{2}{l}{\xspace} \\
        \midrule
        \multicolumn{2}{c}{Panel A: Tabulated at the Census Block Level} \\
        \midrule
                Table Name & Census Block Table Title\\
        \midrule
        P1 & Total Population\\  P6 & Race (6 categories, alone or in combination)\\ P7 & Hispanic or Latino Origin by Race  \\ P8 & Race (63 categories)\\ P9 & Hispanic or Latino, and not Hispanic or Latino by Race \\ P10 & Race for the Population 18 Years and Over \\  P11 & Hispanic or Latino, and not Hispanic or Latino by Race \\ & for the Population 18 Years and Over \\ P12 & Sex by Age  \\ P12A-I & Sex by Age (iterated by Race and Ethnicity)  \\ P14 & Sex by Age for the Population Under 20 Years \\
        \midrule
        \multicolumn{2}{c}{Panel B: Tabulated at the Census Tract Level} \\
        \midrule
                Table Name & Census Tract Table Title\\
        \midrule
        PCT12 & Sex by Age\\ PCT12A-O & Sex by Age (iterated by Race and Ethnicity) \\
        \bottomrule
    \end{tabular}
\flushleft{
{\footnotesize Source: 2010 Summary File 1 technical documentation \citep{sf1:2010:tech}. Table names and titles are taken from this documentation. \\
Notes: Overall, there are 8.6 billion linearly independent statistics in the census block-level tables and 241 million linearly independent statistics in the census tract-level tables. The tables for blocks with zero population are completely zero-filled. The total number of linearly independent statistics counting only blocks and tracts with positive population is 5.0 billion. This table is identical to Table 2 in the supplemental text.}
}
\end{table}

Our experiments use two different subsets of SF1 data as shown in Table \ref{tab:SF1_tables}. The first reconstruction of the HDF, which we denote as \rHDFBT, uses both census block- (\emph{b}) and tract-level (\emph{t}) SF1 tables (those shown in Panels A and B), taking advantage of both the geographic and race detail in the block summaries and the age detail in the tract summaries. The second reconstruction, \rHDFB, uses only the block-level tables (those shown only in Panel A). Thus, the second reconstruction removes the more granular age information found in the tract-level tables while retaining the full race and ethnicity schema used in the block-level data. 

\subsection{Simple example of reconstruction}

To illustrate how our simulated reconstruction attack works in practice, consider \href{https://data.census.gov/table/DECENNIALPL2010.P1?g=1000000US461359664005122}{tables associated with the census block having geocode $461359664005122$} in Yankton, South Dakota. We will use tables taken from the redistricting data and SF1. We begin with \href{https://data.census.gov/table?q=P1&g=1000000US461359664005122}{2010 Redistricting Data, Table P1, identical to 2010 SF1 Table P8} (Race), which indicates that the only resident of this block was a White alone person (i.e., this respondent checked only the ``White'' category on the race question). Next, select table  \href{https://data.census.gov/table?q=P12A&g=1000000US461359664005122}{2010 SF1 Table 12A} (Sex by Age (White alone)), which indicates that this person was 25-29 years old and male. Then select \href{https://data.census.gov/table?q=P7&g=1000000US461359664005122}{Table P7} (Hispanic or Latino Origin by Race (Total Races Tallied)), which indicates that this person reported being  non-Hispanic. So, we have our reconstruction: the microdata record in the confidential HDF of the sole resident of block $461359664005122$ in the 2010 Census is that of a White alone,  non-Hispanic, 25-29 year-old male. 

This case is especially simple due to the small block size,\footnote{In this case, we could have used only SF1 Tables P7 and P12A, without needing redistricting Table P1.} but demonstrates how we can often quickly go from separate tables about census $\{block, sex, age, race$, $ethnicity\}$  to a multivariate description that was present in none of the tables alone. This process is convenient to automate, but, especially in small blocks like $461359664005122$, is often simple to carry out manually. Importantly, this simplicity makes the attack more worrying, not less: it means it is easier for more people to understand and carry out.

For larger blocks, reconstruction can, however, still be achieved. Doing so is harder to easily visualize, and generally requires automation via an appropriate algorithm, although the knowledge and technical requirements for being able to do this are modest for analysts familiar with both the SF1 tables and appropriate software for solving mixed integer linear programs.

\subsection{Basic algorithms used in the reconstruction}

To create record-level images containing the features \{\textit{block, sex, age, race, ethnicity}\} we solve integer programs described in detail in the supplemental text. The algebraic structure of these problems is summarized in supplemental text Equation 4.1, which shows that every published statistic contained in Table \ref{tab:SF1_tables} can be expressed as a linear combination of the records in the HDF. 

To solve these integer-valued linear equations, we convert the algebraic representation to a collection of integer programs. The first reconstruction, \rHDFBT, solves the integer program in Equation 4.4 in the supplemental text using all statistics in Table \ref{tab:SF1_tables}. Thus, \rHDFBT reconstructs the feature set \{\textit{block, sex, age, race, ethnicity}\}. The second reconstruction, \rHDFB, solves the integer program in Equation 4.5 using only the statistics in Panel A of Table \ref{tab:SF1_tables}. Thus, \rHDFB reconstructs the feature set \{\textit{block, sex, agebin, race, ethnicity}\}. The difference between these feature sets is explained in Section \ref{sec:age_in_SF1}. We used the mixed integer linear programming algorithms in Gurobi\textsuperscript{TM} to find the first feasible solution to supplemental text Equations 4.4 or 4.5 in each census tract and block of the U.S. In what follows, we refer to this process as solving the \textit{reconstruction algorithms}.

\subsection{The treatment of age in Summary File 1 and its impact on the reconstruction}
\label{sec:age_in_SF1}
2010 SF1 tabulated age differently depending on the specific table and the level of geographic detail. Tables were published at the county, census tract, and census block levels. Every geographic entity supported by the 2010 Census tabulation system can be constructed by aggregating subsets of these three tabular summaries, which form a strict hierarchy \citep[Chapter 2]{sf1:2010:tech}. At the county and census tract level (e.g., Table PCT12), age was tabulated in single years from 0 to 99 years, then binned into the ranges 100-104 years, 105-109 years, and 110 years and over. 

At the census block level in most tables (e.g., Table P12), age was binned into the following ranges:  
0-4;
5-9;
10-14;
15-17;
18-19;
20;
21;
22-24;
25-29;
30-34;
35-39;
40-44;
45-49;
50-54;
55-59;
60-61;
62-64;
65-66;
67-69;
70-74;
75-79;
80-84;
and 85+. 
Also at the block level, Tables P10 and P11 selected only persons age 18 and older. Finally, the block-level table P14 selected only persons age 20 or younger and encoded age in single years. 

Combining the different age binning and universe selection rules applied at the census block level defines the most detailed age schema that these block-level tables, on their own, can support. That schema has 38 age groups: single year of age from 0 to 21, then:
22-24;
25-29;
30-34;
35-39;
40-44;
45-49;
50-54;
55-59;
60-61;
62-64;
65-66;
67-69;
70-74;
75-79;
80-84;
and 85+. 
We use this 38-bin age schema (feature \textit{agebin}) in our assessments of agreement of the reconstructed HDF with the HDF and CEF. We show in Section \ref{sec:disclosure_risk_hdf} that the 38-bin age schema provides sufficient uniqueness for persons at the census block level to enable reconstruction-abetted reidentification. 

It is also important to note that tables P12 and P14, when combined, give the exact number of males and females in each of the 38 age bins in each census block. Therefore, any remaining reconstruction uncertainty must be in the $race \times ethnicity$ distribution within each sex and age bin.\footnote{Throughout this paper and the supplemental text, the notation \{\textit{race, ethnicity}\} refers to the the subset of features defined by the 63-category \textit{race} and 2-category \textit{ethnicity} features. The notation $race \times ethnicity$ distribution refers to the  $63 \times 2$ table using the 63-category \textit{race} feature crossed with the 2-category \textit{ethnicity} feature. The coding schemes are derived from the definitions used to create lines P8 and P9 in Table \ref{tab:SF1_tables}, Panel A.} 

For \emph{any} reconstructed microdata based on at least the tables in Panel A of Table \ref{tab:SF1_tables}, there is only one possible reconstruction on the feature set \{\textit{block, sex, agebin}\}. Therefore our reconstructions are always exact on that feature set. In our matching algorithms, discussed in Sections \ref{subs:agreematch} and \ref{subs:reidmatch}, we distinguish between matches based on exact age (feature \textit{age}) and those based on binned age (feature \textit{agebin}).  

\subsection{Reconstruction solution variability}
\label{sec:solvar_results}

When an attacker performs a reconstruction, an important question is whether the attacker can determine if the reconstruction in a block is unique, or if the attacker can compute an upper bound on how different two distinct reconstructions yielding the same published tabulations can be. We discuss this issue extensively in the supplemental text Sections 4.5 and 6.1. Summarizing here, all reconstruction error is limited to the race and ethnicity variables and the specific \textit{age} within \textit{agebin}. SF1 tables P12 and P14 provide the exact count of males and females in each of the 38 age bins for each census block. Thus an attacker can always create microdata records with the correct sex and binned age values by expanding those tables. In the \textit{agebin} schema, the only remaining uncertainty is which race and ethnicity value to attach to those records. In the tract-level \textit{age} schema, there is also uncertainty about single year of age within age bin except for the population 21 years and under, since each of these age bins contains exactly one age. 

Variability in reconstruction can offer disclosure protection if there are many microdata sets that are each consistent with a set of published tabulations, if those microdata differ greatly, and if an attacker has no way of choosing between them. In early 2019, motivated by concerns similar to those driving \href{https://journals.sagepub.com/doi/10.2478/jos-2023-0017?icid=int.sj-abstract.similar-articles.6}{Muralidhar and Domingo-Ferrer (2023)}'s work, we began investigating the solution variability of our reconstruction of the HDF. We determined that it was possible to construct a mixed linear integer program to answer the question: if $\text{R}_1$ and $\text{R}_2$ are two record-level reconstructions that reproduce the published tabulations, what percentage of their records can differ between the two solutions? Rather than counting the number of feasible solutions, this measure, which we call ``solution variability'' (\emph{solvar}), upper bounds how different any two solutions can be. 

We calculated \textit{solvar} for single census blocks and sets of census blocks. When $solvar=0$, there is provably only one set of reconstructed microdata that aggregate to the published tabulations. The mathematical representation of our solution variability problem and the integer programs we solved can be found in the supplemental text Sections 4.5 and 4.6; see, in particular, the definitions of \textit{solvar} and \textit{cumsolvar} in supplemental text Equations 4.6 and 4.8, and the integer programs in supplemental text Equations 4.9 and 4.10.

Reconstructed 2010 Census microdata have remarkably little solution variability. That is, about one-third of the reconstructions are exact for the block-level feature set \{\textit{block, sex, agebin, race, ethnicity}\}. The remaining records have very little error---none at all for the features \{\textit{block, sex, agebin}\}, and a small amount in the features \{\textit{race, ethnicity}\}. Given these results, there is no real advantage to collecting additional solutions to the reconstruction algorithms. The main conclusions can be drawn from the 97 million records that are solved exactly.

Table \ref{tab:sol_var_quantiles} reports results on the solution variability in our reconstruction experiments. If the solutions are highly variable, then simulation assessments like those of \citet{rocher:et:al:2019, dick:etal:2023} rather than single-reconstruction assessments, as we propose, would provide additional insight. The results show every fifth percentile of the cumulative distribution of \textit{solvar}, which is always assessed using the \{\textit{block, sex, agebin, race, ethnicity}\} feature set. The cumulative distribution displayed in the table is over census blocks, not persons. Thus there are $\blocksUS/20 = 310,351$ blocks in each cell of the cumulative distribution, the population of which is shown in the ``Population ($\times 10^{3}$)'' column. 

\begin{table}[ht]
\caption{Empirical Percentiles for Block-Level Solution Variability}
\label{tab:sol_var_quantiles}
\begin{tabular}{rrrrrrr}
    \multicolumn{7}{l}{\xspace} \\
    Block      &          & Maximum  &             & Cumulative  & Cumulative & Maximum \\
    Percentile & $solvar$ & $solvar$ & Population  & Population  & $solvar$   & Cumulative \\
     (\%)      & (\%)     &  (\%)    & ($\times 10^{3}$) & ($\times 10^{3}$) &  (\%)      & $solvar$ (\%)  \\ 
    \midrule
5&0.0&0.0&6,398&6,398&0.0&0.0 \\ 
10&0.0&0.0&6,376&12,774&0.0&0.0 \\ 
15&0.0&0.0&6,381&19,155&0.0&0.0 \\ 
20&0.0&0.0&6,372&25,527&0.0&0.0 \\ 
25&0.0&0.0&6,391&31,918&0.0&0.0 \\ 
30&0.0&0.0&6,376&38,294&0.0&0.0 \\ 
35&0.0&0.0&6,376&44,670&0.0&0.0 \\ 
40&0.0&0.0&6,376&51,046&0.0&0.0 \\ 
45&0.0&0.0&6,381&57,427&0.0&0.0 \\ 
50&0.0&0.0&6,372&63,799&0.0&0.0 \\ 
55&0.0&0.0&6,412&70,211&0.0&0.0 \\ 
60&0.0&0.0&6,376&76,587&0.0&0.0 \\ 
65&0.0&0.0&6,380&82,967&0.0&0.0 \\ 
70&0.0&0.0&14,271&97,238&0.1&0.2 \\ 
75&1.7&3.4&34,272&131,510&0.9&1.8 \\ 
80&4.7&9.3&28,281&159,790&1.8&3.6 \\ 
85&7.3&14.6&28,776&188,566&2.9&5.8 \\ 
90&10.5&21.1&30,319&218,884&4.2&8.4 \\ 
95&14.6&29.3&36,466&255,351&6.1&12.3 \\ 
100&21.1&42.1&53,395&308,746&10.0&20.1 \\ 
\bottomrule
\end{tabular}
\flushleft
{\footnotesize Notes: This table is based entirely on public data. All statistics are displayed to full precision. Solution variability is the statistic $solvar$ in Equation 4.10 in the supplemental text, which has been sorted in increasing order by census block. Maximum $solvar$ is an upper bound on the solution variability of any reconstruction as given in Equation 4.7 in the supplemental text. Cumulative $solvar$ is the statistic $cumsolvar$ in Equation 4.8 in the supplemental text when the blocks have been sorted in increasing order of $solvar$. Maximum cumulative $solvar$ is an upper bound on the cumulative solution variability of any solution to the \rHDFBT reconstruction. Percentiles are defined over blocks, not persons. Block ties in the definition of $solvar$ percentiles were broken randomly. Consequently, running the replication package for this table may result in minor variations in the population, cumulative population, and cumulative solution variability columns. Cumulative $solvar$ is subject to accumulated tiny fractional values of $solvar$, which is why it is 0.1 for the 70th percentile and not 0. This table is identical to Table 6 in the supplemental text.
}
\end{table}

For 70\% of all blocks, representing 97,238,000 person records, solution variability is 0. Furthermore, in any other reconstruction solution for \rHDFBT no more than 20.1\% of all records can differ from their HDF record on the value of even a single feature, evaluated on the \{\textit{block, sex, agebin, race, ethnicity}\} schema. The features \{\textit{block, sex, agebin}\} can never show any solution variability, and all additional \textit{age} variability in solutions for \rHDFBT must be within $agebin$ categories.  The reconstructed records for those individuals with  zero solution variability are thus guaranteed to match their confidential HDF records exactly using the \textit{agebin} schema. 

The maximum solution variability for any block in the $75^{th}$ quantile is 3.4\%. Since there are 34,272,000 people in this quantile, this means that at most 1,165,000 records in this quantile can differ from their confidential source on \textit{race} or \textit{ethnicity}. The maximum cumulative solution variability given any feasible solution for all census blocks, containing all 308,746,000 persons, is just 20.1\%. 
Thus, if \textit{agebin} is sufficient for high agreement and reconstruction precision, the additional solution variability of the feature \textit{age} within the feature \textit{agebin} does not matter.  The detailed solution variability results can be found in the supplemental text Section 6.1.

Finally, we note that \citet{hawes:2022,abowd:hawes:2023} report reconstruction agreement and solution variability results based on 32 of the 34 tables shown in Table \ref{tab:SF1_tables}---excluding P8 and P10.\footnote{These results were the basis for presentations the Census Bureau's Scientific Advisory Committee, whose working group on differential privacy was monitoring the research and implementation decisions underlying the 2020 Census Disclosure Avoidance System.} When those two tables are omitted, solution variability is zero for only 65\% of blocks (82.9 million persons), demonstrating empirically the mathematical property of the solution variability---as more tables are added the solutions become less variable. In this case, two tables with additional \textit{race} data for all persons and persons age 18 and older increased the number of zero solution variability records by 14.3 million. That is 14.3 million additional people for whom the attacker is certain that the reconstructed record matches the HDF record exactly using the \textit{agebin} schema. 

\subsection{Extended example of reconstruction in a single census block}

\begin{table}[ht]
    \centering
    \caption{The exact reconstruction of census block 1001, in census tract 51.03, in Jefferson County, Alabama (01073)}
    \resizebox{15cm}{!}{
    \begin{tabular}{|p{8mm}|p{26mm}|p{5mm}|p{6mm}|p{11mm}|p{9mm}|p{9mm}|p{9mm}|p{9mm}|p{9mm}|p{7mm}|p{8mm}|p{5mm}|}
 \hline
 Row ID &	Block ID & Sex & Age & Binned Age & White & Black & AIAN & Asian & NHPI & SOR & Hisp. & ...\\
        \midrule
\hline
20 & 010730051031001 & M & 1 & 1 & N & Y & N & N & N & N & N & ...\\
\hline
21 & 010730051031001 & M & 17 & 17 & N & Y & N & N & N & N & N & ...\\
\hline
22 & 010730051031001 & M & 32 & 30 - 34 & N & Y & N & N & N & N & N & ...\\
\hline
23 & 010730051031001 & M & 46 & 45 - 49 & N & Y & N & N & N & N & N & ...\\
\hline
24 & 010730051031001 & M & 59 & 55 - 59 & N & Y & N & N & N & N & N & ...\\
\hline
25 & 010730051031001 & M & 68 & 67 - 69 & Y & N & N & N & N & N & N & ...\\
\hline
26 & 010730051031001 & F & 18 & 18 & N & Y & N & N & N & N & N & ...\\
\hline
27 & 010730051031001 & F & 33 & 30 - 34 & N & Y & N & N & N & N & N & ...\\
\hline
28 & 010730051031001 & F & 37 & 35 - 39 & N & Y & N & N & N & N & N & ...\\
\hline
29 & 010730051031001 & F & 45 & 45 - 49 & N & Y & N & N & N & N & N & ...\\
\hline
30 & 010730051031001 & F & 58 & 55 - 59 & Y & N & N & N & N & N & N & ...\\
\hline
31 & 010730051031001 & F & 58 & 55 - 59 & N & Y & N & N & N & N & N & ...\\
\hline
32 & 010730051031001 & F & 59 & 55 - 59 & N & Y & N & N & N & N & N & ...\\
\hline
33 & 010730051031001 & F & 71 & 70 - 74 & Y & N & N & N & N & N & N & ...\\
\hline
34 & 010730051031001 & F & 74 & 70 - 74 & Y & N & N & N & N & N & N & ...\\
\hline
35 & 010730051031001 & F & 77 & 75 - 79 & N & Y & N & N & N & N & N & ...\\
\hline
    \end{tabular}}

    \label{tab:Example_1}
    \flushleft\footnotesize 
    Source: Rows 20 to 35 of rhdf\_bt\_0solvar\_extract.xlsx in the replication archive (See the Data availability section). \\
    Notes: ``M'' is ``male'' and ``F'' is ``female''; ``AIAN'' is ``American Indian and Alaska Native''; ``NHPI'' is ``Native Hawaiian and Pacific Islander''; ``SOR'' is ``Some Other Race''; and ``Hisp.'' is ``Hispanic or Latino''.
\end{table}

This example of solving the reconstruction algorithms demonstrates that there is a unique set of 16 records, displayed in Table \ref{tab:Example_1}, that adhere to the constraints imposed by the tabular releases for this particular block. The existence of a unique solution for these 16 records confirms an exact reconstruction of this census block without using any confidential data. To produce these block-level records, the reconstruction algorithms use the following information from the 2010 SF1. From Table P1, the reconstruction algorithm produces exactly 16 records for this block. Their features are determined as follows. From Table P12, there are 1 male age 30-34, 1 male age 45-49, 1 male age 55-59, 1 male age 67-69, 1 female age 30-34, 1 female age 35-39, 1 female age 45-49, 3 females age 55-59, 2 females age 70-74, and 1 female age 75-79. From Table P14, there are 1 male age 1, 1 male age 17, 1 female age 18. From Table P9 (redistricting table P2),  there are 0 Hispanics. From Table P8 (redistricting table P1), there are 4 Whites, 12 Blacks, and 0 in all other categories of race. From Table P12I, there are 1 White male age 67-69, 1 White female age 55-59, and 2 White females age 70-74. From Table P12B, there are 1 Black male age 0-4, 1 Black male age 15-17, 1 Black male age 30-34, 1 Black male age 45-49, 1 Black male age 55-59, 1 Black female age 18-19, 1 Black female age 30-34, 1 Black female age 35-39, 1 Black female age 45-49, 2 Black females age 55-59, and 1 Black female age 75-79. The solution variability algorithm establishes that there is exactly one set of 16 records that satisfy these constraints.

An attacker has enough information to find unique persons in this census block without using any confidential data. The attacker who knows name, address, sex and age learns with certainty that the name and address associated with RowID 25 represents a non-Hispanic White person on the confidential 2010 Census record used to produce this table. Such knowledge is only possible because RowID 25 was included in the input file used to produce SF1; that is, only because the response of the person associated with RowID 25 was used to make the tables in SF1. Such an inference is not statistical; it is disclosive. There is no uncertainty about the content of this particular record in the confidential 2010 Census HDF and no uncertainty about its association with the attacker’s data. The rows in Table \ref{tab:Example_1} are known with certainty to be in the confidential HDF, thus proving that aggregation into tabular summaries, as implemented for the 2010 Census, is insufficient to protect the confidentiality of the 2010 HDF. 

When considering the example above, note that while some data subjects may not consider their response information about race or ethnicity as particularly in need of confidentiality protection, we stress that race and ethnicity are used here for illustration purposes, not because of any special structure these variables have that is useful for a reconstruction or reconstruction-abetted reidentification attack. Presumably, many more respondents would be concerned about the confidentiality of other attributes that can be revealed via reconstruction such as whether they were in a same-sex marriage, whether their child was adopted, if they were an older person living alone, if they have exceeded their lease's occupancy limit, etc. 

The existence of this exact reconstruction of a complete census block of records in the confidential input data and the fact that no confidential data are required to confirm the correctness of these records is the failure of the 2010 Census disclosure avoidance system that relied on differing SDL methods for tabular vs. microdata products. 
At a larger scale, the magnitude of this problem for the 2010 Census becomes even more readily apparent. 

\subsection{Implications of successful reconstruction}

Our reconstruction results with zero or very limited solution variability imply that for much of the U.S. population the record-level image of the features used to create the census tract- and block-level data shown in Table \ref{tab:SF1_tables} is essentially an exact copy of the confidential HDF on the binned-age schema. Thus, the SF1 data shown in Table \ref{tab:SF1_tables} are equivalent to the confidential microdata HDF records for the \{\textit{block, sex, agebin, race, ethnicity}\} feature set. The release of these microdata was prohibited by the 2010 Census disclosure avoidance rules \citep{mckenna:2019:microdata}. 

The ability to successfully reconstruct the HDF from this subset of 2010 Census published tabulations, and a potential attacker's confidence in the accuracy of that reconstruction, because solution variability can also be calculated without using the confidential data, demonstrate that aggregation as implemented for the 2010 Census tabular products was not a sufficient SDL strategy to protect the confidential HDF. Full statistical details of the reconstruction accuracy can be found in the supplemental text Section 6.2.

Permission to release the complete reconstructed HDF was approved in 2022 under clearance number CBDRB-FY22-DSEP-004. The public replication archive for this paper includes a sample of the reconstructed records in 29 tracts with  zero solution variability, all necessary inputs and programs to reconstruct the entire \USpopulation person records for \rHDFBT and \rHDFB from public data, and everything required to confirm our solution variability results. This portion of the replication archive does not use any confidential data. Further information can be found in the Data availability section at the end of this paper.

\section{Can one confidently and accurately reidentify specific individuals in microdata records reconstructed from the 2010 Census tables?}
\label{sec:disclosure_risk_hdf}

Having demonstrated that aggregation into tabular summaries was insufficient to protect against the accurate reconstruction of the 2010 Census HDF, we now turn to the larger question of whether the Census Bureau was correct to assume that the HDF microdata posed a substantial disclosure risk warranting additional SDL methods beyond the record swapping already present in the file. More specifically, we examine this issue from the targeted person's perspective, asking whether the reconstructed HDF records from Section \ref{sec:aggregation_hdf} themselves pose such a disclosure risk because they enable highly accurate inferences about the target person that would not have been possible in the absence of that person's actual census responses.

To demonstrate this vulnerability, we must be explicit about the attacker's data requirements and how we measure their quality. Reidentification is accomplished via record linkage between the attacker's database and the reconstructed HDF microdata. The record linkage is based on features that are common to both the attacker's database and the HDF microdata. These common features are called linking variables or quasi-identifiers. We use only four linking variables: name, address, sex, and birth date. 

The quality of the attacker's database is measured by the agreement between the values of its linking variables and the corresponding values in the actual 2010 Census responses. In higher-quality attacking data, the name, address, sex, and birth date would be the same as the values reported by the census respondent relative to the reference date of April 1, 2010. In lower-quality attacking data some or all of these features would differ from the values the respondent reported at the time of the 2010 Census. In particular, the name might be recorded differently, the birth date might be incomplete, the address might be relative to a different reference date, or some of the linking variables might be missing in a lower-quality attacker database. Finally, while we use commercial data to model one of our attacker databases, the attacker could be a person, business, or government entity with their own proprietary or administrative record datasets. Those attacker databases would be constructed from the name, address, sex, and birth date information in their systems.

\subsection{Additional data sources used in the reidentification studies}
\label{sec:reid_sources}
In addition to the reconstructions \rHDFBT and \rHDFB (the inputs for which are described in Section \ref{sec:inputs_recon}), the reidentification portion of the study used two additional data sources. 

One attacker file (COMRCL) is a collection of commercially available data from the 2010 time frame meant to approximate external information that could have been easily purchased by an attacker at the time of the 2010 Census. We created the COMRCL file used for our reidentification experiments by combining data extracts originally purchased from four commercial providers between 2009 and 2011 in support of the 2010 Census evaluations.\footnote{The four commercial databases were provided by Experian Marketing Solutions Incorporated, Infogroup Incorporated, Targus Information Corporation, and VSGI LLC. The databases used are the same as in \citet{rastogi2012} except that we excluded data provided by the Melissa Data Corporation, which contain address information but not sex and age data.} We show in section \ref{sec:attacker_files} that the COMRCL data serve as the background knowledge of an attacker with lower-quality information contemporaneous with the release of the 2010 SF1 because there is relatively poor agreement between the COMRCL linking variables and the values in the confidential census records. 

While the database schema and the purposes for which these commercial databases were originally collected vary, they all share certain attributes. All have basic personal identifying information (PII) including names, addresses, sex, and birth dates from which we construct the linking variables. The vintage 2010 versions of these databases that we used did not include self-reported race and ethnicity data.\footnote{Race and ethnicity data are modeled in some of the commercial databases \citep{rastogi2012}, but we did not use this information.}

The reidentification studies also used the confidential 2010 Census response information from the CEF, introduced in Section \ref{subs:SDL_2010}. The CEF is the archival, unswapped precurser to the HDF discussed in Section \ref{sec:aggregation_hdf}. We used the CEF to confirm the accuracy of suspected reidentifications. 

We also used an extract of the CEF ($\text{CEF}_{atkr}$), described more fully in Section \ref{sec:reid_high}, to create a second attacker file that simulates how well an attacker might perform using higher-quality external information. Because we base the linking variables on the actual census responses, the $\text{CEF}_{atkr}$ attacker is guaranteed to find a link in the confidential census data, but not necessarily in the reconstructed HDF.

\subsection{Record linkage and person/household identifiers at the Census Bureau}
\label{sec:identifiers}

Our reidentification studies must use a record-linkage system that standardizes names and addresses so that persons in one source (e.g., COMRCL) can be associated with unique persons in other sources (e.g., CEF). As part of the Census Bureau's internal confidentiality safeguards, the respondent's name and address are not stored in the HDF or the CEF. Rather, census data processing links the household's physical address to an identifier called the Master Address File Identifier (MAFID) and links the person's name to a person identifier called a Protected Identification Key (\textit{pik})  

Not all records in the CEF have a \textit{pik}, and in some cases the same \textit{pik} appears on multiple records. For the purposes of this paper, we refer to the subset of records in the CEF with a distinct \textit{pik} within the record's census block as the \emph{data-defined population}. Record-linkage systems work with data-defined populations. When there is insufficient data in a person's response, statistical methods (e.g., imputation) are used for tabulations, but these methods inherently confound record-linkage systems. Hence, the focus on the data-defined population. 

To create the data-defined population, we removed all CEF records with insufficient data to assign a \textit{pik}. Then, if there were multiple records with the same \textit{pik} within a census block, one record was randomly chosen. The data-defined population is thus 276,000,000 records ($89\%$ of all records in the CEF). Records with duplicate \textit{pik}s within a single census block appeared in 15\% of blocks. In total, 1\% of records with a \textit{pik} were removed by this unduplication. The remainder of the difference between the total 2010 Census population and the data-defined population is incomplete or imputed census responses that did not receive a \textit{pik}. 

Additional details regarding the internal 2010 Census data files and record-linkage procedures can be found in the supplemental text Sections 3.1 through 3.4. Supplemental  text Table 1 contains a complete inventory of the input and output data files used in this study along with a catalog of the features found in each file.

\subsection{Circa 2010 commercial ``attacker'' files}
\label{sec:attacker_files}

We harmonized the feature sets for the commercial data to match the schemas used in the CEF, as indicated in the COMRCL row of supplemental text Table 1. These data were originally acquired because the features we use---name, address, sex, and birth date in particular---were expected to closely resemble those collected on the 2010 Census. In our harmonization, name and address were mapped to \textit{pik} and MAFID, respectively.  

Table \ref{tab:cef_comrcl_match} shows that among the COMRCL records only 106,300,000 ($37.1\%$) matched a CEF record on $\lbrace$\textit{pik, block, sex, agebin}$\rbrace$, i.e., using binned age instead of exact age. Thus, the commercial data used here are not very accurate compared to the 2010 CEF, and we do not rule out the possibility that better quality data may have been available in 2010.\footnote{In particular, the Census Bureau did not purchase the 2010 version of the LexisNexis Public Record Search database, another commercial product similar to the ones we used in COMRCL, because it could only be used with its own proprietary software.} Better external data were available for the 2020 Census \citep{brown:etal:2023}, including both commercial and administrative-record sources.

\begin{table}[ht]
\caption{Overlap of Data-Defined Person Records in  CEF and COMRCL Databases}
\label{tab:cef_comrcl_match}
\centering
\begin{tabular}{l|rr|r|r}
    \multicolumn{5}{l}{\xspace} \\
    &  \multicolumn{2}{c|}{In CEF Universe}   & Not in CEF  &  \\ 
    &  Matched    &  Unmatched                & Universe    & Total \\ 
    & ($\times 10^{3}$) & ($\times 10^{3}$)               & ($\times 10^{3}$) & ($\times 10^{3}$) \\
    \midrule 
    Records in COMRCL&106,300&180,400&2,449&289,100 \\ 
    Records not in COMRCL&169,700&&& \\ 
    \midrule
    Total & 276,000&&& \\ 
    \bottomrule
\end{tabular}
\flushleft
{\footnotesize Notes:  Counts rounded to four significant digits to conform to disclosure limitation requirements. The commercial data contain census block geocodes not found in the CEF universe. The columns ``In CEF Universe, Matched'' and ``In CEF Universe, Unmatched'' reflect only records with 2010 census block geocodes in the CEF universe. ``Matched'' means the records agree on the feature set \{\textit{block, pik, sex, agebin}\}. The balance of the data-defined COMRCL records are shown in the column ``Not in CEF Universe.'' The research in this paper can only use those records in the CEF universe (106,300 + 180,400 = 286,700 thousand). When COMRCL records are classified as modal or nonmodal we use matched records for which CEF attributes are known. In this case ``matched'' means the records agree on \{\textit{pik, block}\} only. A trivial number of COMRCL records that match on these two variables disagree on \{\textit{sex, agebin}\}. Hence, the universe for analyses that distinguish modal and nonmodal COMRCL records is also 106,300 thousand. This table is identical to Table 3 in the supplemental text.}
\end{table}

\subsection{High-level reidentification methodology}
\label{sec:reid_high}

At a high level, we sought to determine whether the HDF microdata constitutes a disclosure risk by simulating a record-linkage based reidentifcation attack using the reconstructed HDF to see how accurately an attacker can infer race and ethnicity from the reconstructed microdata. Thus, we link \rHDFBT and \rHDFB separately to both COMRCL (to simulate an attack using lower-quality external attacker data) and $\text{CEF}_{atkr}$ (to simulate an attack using higher-quality external attacker data). $\text{CEF}_{atkr}$ is a specially constructed extract from the CEF that includes the linking variables (quasi-identifiers) $\{block, sex, age\}$ and the person identifier \textit{pik} but no other variables---specifically, not \textit{race} or \textit{ethnicity}.

By record linkage of the quasi-identifiers \{\textit{block, sex, (age} or \textit{agebin)}\} in the reconstructed microdata to attacker databases that include names (feature \textit{pik}), we create \emph{putative reidentifications}. To enhance the attacker's database, we attach the data on \{\textit{race, ethnicity}\} found on \rHDFBT or \rHDFB to the \{\textit{pik, block, sex, (age} or \textit{agebin)}\} information in the attacker's database---either COMRCL or $\text{CEF}_{atkr}$. 
Finally, to evaluate the accuracy of the putative reidentifications, and classify \emph{confirmed reidentifications}, we match putative reidentifications to the full CEF, linking on \{\textit{pik, block, sex, (age} or \textit{agebin)}\} and comparing the \{\textit{race, ethnicity}\} inferred from the reconstructed microdata (attached to a putatively reidentified person) to that person's actual census responses in the CEF. We label the reidentification \emph{confirmed}, when \{\textit{pik, block, sex, (age} or \textit{agebin), race, ethnicity}\} all match in either the  \textit{age} or \textit{agebin} schema. 
Finally, we use the ratio of confirmed to putative identifications as the measure of precision or accuracy for the attack.

To make the meaning of this exercise as clear as possible, we restate the attacker model we are simulating concisely here. The attacker is an entity external to the Census Bureau with no access to the confidential data contained in HDF or CEF. The attacker has access to all published 2010 Census data---every table in SF1, in particular. The attacker selects a subset (possibly the universe) of these tables and performs record-level reconstruction, possibly using the algorithms in this paper (that is, the algorithms described in the supplemental text Sections 4 and 5). The attacker also has access to an external database that contains name and \textit{address} (or some other personal identifying information sufficient to tag a unique person like Social Security Number) and quasi-identifiers---features that match some of the features in SF1---specifically, census \textit{block} (geocoded from \textit{address}), \textit{sex}, and \textit{age}. 

Notice that we have explicitly used the same feature set definitions for the attacker's database as we used in our reconstruction. This is an essential characteristic of a record-linkage attack---the attacker knows enough about the definitions of the feature set for SF1 to construct quasi-identifiers with the same schema. The attacker matches the external data to the reconstructed record-level data from SF1 using the matching variables (quasi-identifiers) \{\textit{block, sex, age}\} and adds the other variables reconstructed from SF1 to the external database. These added variables can include any feature tabulated in SF1, including those tabulated for housing units and householders because those data can be reconstructed as part of the record for ``person 1'' or ``householder,'' the individual who completed the census responses for the persons living in the housing unit. In our experiments the extra variables are \{\textit{race, ethnicity}\}.

The details of our matching and reidentification methods and specific algorithms can be found in the supplemental text Section 5. Implementing additional features in the reconstructed data is outside the scope of this paper; however, as an illustration, we sketch the details for adding the feature ``relationship to householder'' in the supplemental text Section 5. 

At this point the attacker is using the information from what we call putative reidentifications as record-level variables on the external database. The attacker may not care about the quality of these putative reidentifications; however, because the solution variability in the schema of our attack is zero for 70\% of all census blocks using just the 34 tables in Table \ref{tab:SF1_tables}, the attacker may be satisfied with the 97 million persons for whom they know with certainty that they have exactly the information on the confidential internal HDF. Alternatively, the attacker can keep adding tables from SF1 until solution variability is zero for as large a portion of the target population as desired. For all records for which solution variability is 0, the attacker knows with certainty that the data added from SF1 for those persons exactly match the confidential  HDF records.

In our experiments, we use the CEF itself as a labeled database to confirm the accuracy of inferences about \{\textit{race, ethnicity}\}. An external attacker would need a similar labeled database to confirm the accuracy of those inferences. It need not be a complete enumeration like the CEF. Small-scale sample surveys or methods like those in \citet{rocher:et:al:2019} and \citet{dick:etal:2023} could be used. Or the data steward---the agency responsible for decisions about data curation and publication---could allow the publication of data about how accurate those inferences would be, as the Census Bureau is doing by releasing the statistics found in this paper. If those inferences are sufficiently more accurate than well-specified baselines, then the confidentiality protections have failed. Such failure is not a \{0,1\} event. It is a continuum on $[0,1]$. Our methodology calibrates this continuum using the precision of inferences about the features added to the external data from SF1 via the record-linkage attack---specifically, the ratio of confirmed to putative reidentifications. 

Our vulnerable population (those most likely at risk for confidentiality-breaching reidentification) is persons who are unique on the linking variables (\textit{block, sex, agebin}) and also differ from their neighbors on the attributes of interest (\textit{race, ethnicity}). We use the census block to define their neighbors.  
Our methodology properly distinguishes between vulnerable populations---those where baseline statistical models have low precision and fail to make correct inferences---and nonvulnerable populations---those where baseline statistical models have high precision and generally make correct inferences. We use the results for vulnerable populations to illustrate the confidentiality protection failures in the 2010 Census publications. Specifically, the high reidentification precision rates for vulnerable populations are entirely due to the inclusion (and exposure) of the vulnerable persons' data in the published tables. 

The most vulnerable persons are population uniques. Delete their data, and the tables that would have contained their responses are completely silent on the risky feature values; that is, the precision of baseline statistical inferences is exactly 0.\footnote{The precision of the statistical baseline is exactly zero if the inference is based on the modal value of $\{race, ethnicity\}$ or the observed proportions in the block.} The vulnerable populations were supposed to be protected by the use of record-level swapping targeting specific population-unique households, but those protections did not recognize how extensive and widespread such vulnerable populations were nor how their data could be reconstructed from the ensemble of published tables. 

\subsection{Agreement match} 
\label{subs:agreematch}

For the reidentification attack, our matching algorithm is straightforward. Given two databases and a set of common features, 
the algorithm matches records on the set of features exactly and without replacement. It iterates over the rows in the left database (the reconstructed HDF records) ($L$) searching, in order, over the rows in the right database, the external attacker file ($R$) to look for the first (if any) record that matches on all the selected features. If a matching record is found, the matching records in the left and right databases are both removed and the algorithm continues to the next record in the left database, again looking for a match in the right database. Notice that an essential feature of this matching algorithm is that every record in the left and right databases is at risk for one, and only one, match. It is not possible for a common record type in the right database to be linked to many records in the left database. Failure to enforce this condition results in spurious claims about match rates as, for example, in \citet{ruggles:vanriper:2022}.

The \rHDFBT, \rHDFB, CEF, and HDF have an overlapping feature set that supports the $\lbrace$\textit{block, sex, age, race, ethnicity}$\rbrace$ schema, as well as the associated schema where $age$ is replaced by $agebin$. In order to measure how well reconstructed records match the underlying confidential data in HDF and CEF, we match the reconstructed microdata to our confidential databases on common features. The algorithm works block-by-block in two passes. First, it finds all matches using  $\lbrace$\textit{block, sex, age, race, ethnicity}$\rbrace$. Next, it finds any remaining matches using $\lbrace$\textit{block, sex, agebin, race, ethnicity}$\rbrace$. The algorithm returns the unmatched records, counts of the matched records in both passes, and indices of the matched records in the original databases for both passes by census block. Additional details and the exact Algorithm 1 (match function) and Algorithm 2 (agreement function) can be found in the supplemental text Section 5.1.

\subsection{Reidentification match} 
\label{subs:reidmatch}

Our reconstruction-abetted reidentification attack uses the common features between the reconstructed database and the attacker database (either COMRCL or $\text{CEF}_{atkr}$) to attach features previously unknown to the attacker, in this case $\lbrace$\textit{race, ethnicity}$\rbrace$, to the attacker database by linking on the common features \{\textit{block, sex, (age} or \textit{agebin)}\}. Thus, the attacker learns information about the database members that was previously not available. 

To evaluate  the strength of the inference an attacker might achieve from access to improved auxiliary data, we compare the results from the lower-quality commercial data that were acquired by the Census Bureau contemporaneously with the 2010 Census with higher-quality attacker database formed by extracting \{\textit{pik, block, sex, age}\} directly from the CEF, called $\text{CEF}_{atkr}$. When $\text{CEF}_{atkr}$ is the attacker database, we exclude $pik$ from the putative match linkage, using only the linking features \{\textit{block, sex, (age} or \textit{agebin)}\}, as we do with the commercial data. In general, we denote the attacker's external database as $D_X$ and the reconstructed database as $D_R$. Note that $D_X$ may have incomplete coverage, $rows(D_X) < rows(D_R)$ and may contain incorrect information relative to the confidential data. Additional details on the creation of $\text{CEF}_{atkr}$ can be found in the supplemental text Table 1 and Section 5.

A successful match between records in $D_R$ and $D_X$, based on the common features $\lbrace$\textit{block, sex, (age} or \textit{agebin)}$\rbrace$, is called a  putative reidentification, since the attacker must collect additional information, possibly through independent field work or simulation studies, to verify that the putative match is correct.\footnote{If the process of classifying the putative reidentifications in the reconstructed data as ``correctly matched'' or ``incorrectly unmatched'' to the external database is considered a statistical classifier, then the attacker needs a labeled training sample.} Like the agreement match algorithm, our record-linkage algorithm consists of two passes that first match on \textit{age} for the complete input databases and then match on \textit{agebin} for the unmatched residual from pass one. The algorithm returns an enhanced attacker external database $D_{X+}$ consisting of records from $D_X$ for which a match was found in the reconstructed database with sensitive features $\lbrace$\emph{race, ethnicity}$\rbrace$ appended. Additional details of the putative match and the exact Algorithm 3 can be found in the supplemental text Section 5.2.

Given the enhanced attacker external database $D_{X+}$, we next determine if the $\lbrace$\textit{race, ethnicity}$\rbrace$ values appended from the reconstructed data match the confidential census responses in the CEF exactly. Like the agreement and putative reidentification algorithms, our confirmation algorithm consists of two passes that first match on \textit{age} for the complete input databases and then match on \textit{agebin} for the unmatched residual from pass one. Records that meet the matching criteria are called confirmed reidentifications. Additional details of the reidentification match and the exact Algorithm 4 can be found in the supplemental text Section 5.2.

\subsection{Assessing confidentiality-violating inferences}
\label{subs:baselines}

A common misconception about SDL techniques is that their objective is to prevent an attacker from revealing or confidently inferring something about a particular data subject from the published statistics. Another misconception is that the objective of SDL is to prevent an attacker from using those published statistics to infer something about a data subject that may be reflected in the subject's confidential census response. The correct objective of SDL is more nuanced---it is to prevent an attacker from using the published statistics to infer something about the data subject that they would not have otherwise been able to infer if the subject's record were not used to produce the published statistics \citep{Dwork2008}. We call this ``leave-one-out'' reasoning and explain its origins and application to SDL in supplemental text Sections 1, 2.4 and 2.5. 

For example, according to the 2020 Census, the population of Lincoln County, WV was 20,463 people, 96.7\% of whom identified as White alone. Armed with this information, an attacker could easily infer (with 96.7\% confidence) that a particular resident of Lincoln County (we will call her ``Jane Doe'' for the purposes of this example) is White, and that Jane's 2020 Census response would reflect this. This type of high confidence inference is neither a violation of statistical confidentiality nor a failure of SDL to protect Jane's census record. It is a statistical or scientific inference made possible by the overall homogeneity of the community and the attacker's knowledge that Jane is a member of that community. 

The same inference about Jane's race could still be made, with the same degree of confidence, even if Jane had not responded to the census. The attacker's inference, in this example, is made purely from generalizable knowledge about the characteristics of the community in which Jane lives, and is not due to any leakage of Jane's unique census responses through the published statistics. Accurate assessment of the success or failure of an SDL mechanism, therefore, must distinguish between confidentiality-violating inferences (which SDL seeks to protect against), and statistical or generalizable inferences (which SDL does not, and perhaps more importantly should not, aim to preclude).\footnote{This is precisely the error noted in \citet[Supplement p. 8]{jarmin:etal:2023} regarding the analyses in \citet{kenny:et:al:2021}.}

In order to differentiate confidentiality-violating inferences from generalizable ones, the results of the reidentification attack must be compared to inferences that would be possible in a confidentiality-preserving counterfactual setting in which the same statistics are published, except that a target individual's record has been removed. In this ideal case, we would compare inferences made about the target person from the published 2010 Census data to inferences that would be made in the counterfactual world in which the target person was removed. Exact comparisons of this sort are computationally intractable because they involve removing a target individual from the full microdata file, re-swapping and re-tabulating the data, performing reconstruction and reidentification to make inferences about the individual, and then repeating this process for every individual in the United States, or at least a meaningful subset.

\subsubsection{Statistical baselines} To make the assessment of confidentiality-violating inferences feasible, we instead develop baselines that focus on small populations (census blocks) and focus our attention on inferences about persons who do not match the modal race and ethnicity in their block. The intuition is that persons with nonmodal $\lbrace$\textit{race, ethnicity}$\rbrace$ and who are unique on $\{block,sex, age\}$ or $\{block, sex,agebin\}$, according to the operative schema, are most at risk for confidentiality-violating inferences due to the leakage of their census responses through the published statistics, rather than through statistical or generalizable inferences about the characteristics of the communities around them.

For these individuals, the reconstruction-abetted reidentification attack could not have even identified a corresponding record as a putative reidentification had the target record been absent from the CEF. Such cases are not uncommon---fully $44\%$ of all persons in the CEF are unique on $\{block, sex, age$ or $agebin\}$. To compare against what an attacker could learn purely through generalizable inference, we posit two ``statistical guessing'' baselines that generate inferences by using either the modal $\lbrace$\textit{race, ethnicity}$\rbrace$ in a block (MDG, for modal guesser), or by guessing with probabilities proportional to the frequency of each $\lbrace$\textit{race, ethnicity}$\rbrace$ in a block (PRG, for probability guesser). 

We compare the performance of these statistical baselines to the performance achieved by the reconstruction-abetted reidentification attack. We assert that if an attacker can make better inferences about these non-modal individuals as a result of the reconstruction-abetted reidentification than the attacker could otherwise make using either of these statistical guessing models, it can only be because information from the individuals' unique census responses is being revealed through the published statistics in a confidentiality-violating manner.\footnote{By considering only a small subset of possible inferences and by allowing these statistical guessers to use information that implicitly treats the release of block-level $\textit{race} \times \textit{ethnicity}$ tables even in very low-population blocks as statistical rather than confidentiality-eroding, this approach probably underestimates the true extent of confidentiality violations. However, it is computationally tractable and identifies a class of inferences and a group of target persons for which it is difficult to view the resulting inference precision as anything but a confidentiality violation.} Additional details and the exact definitions of our statistical baselines can be found in the supplemental text Section 5.3.

We illustrate these statistical guessing baselines with an example. Given the homogeneity of race and ethnicity within blocks, it would be reasonable to use the $\lbrace race$, $ethnicity \rbrace$ data from other individuals in a block to make inferences about the target person. Suppose a block consisted of 10 people $\lbrace r_1,\dots, r_{10} \rbrace$, with the first 9 being White alone, and the 10\textsuperscript{th} person being Asian alone. All are non-Hispanic. When the target person is $r_1$, the attacker in the counterfactual world ($r_1$'s record is removed) would see 8 White alone individuals and 1 Asian alone. MDG would predict that the target person is White while PRG would guess randomly in proportion to each type, assigning White alone with probability 8/9 and Asian alone with probability 1/9. Alternatively, if the target person is $r_{10}$, the attacker in the counterfactual world would see 9 White alone individuals, and both the modal and proportional guessers would incorrectly guess White alone with certainty. Repeating such an exercise across all individuals would result in a modal guesser achieving an accuracy of 90\% (the only mistake coming when the target individual is Asian alone) while the expected accuracy of the proportional guesser is approximately 81.1\%. To simplify the calculations of these baselines, we use  upper bounds. An upper bound on the accuracy of the modal guesser is the fraction of the block's population that reports the modal $\lbrace$\textit{race, ethnicity}$\rbrace$ in the block. The upper bound on the accuracy of the proportional guesser is $\sum_i p_i^2$, where $\lbrace p_1,p_2,\dots \rbrace$ are the proportions of the block's population of each of the 126 $race \times ethnicity$ cells.

Note that the modal guesser is targeting overall accuracy and would perform poorly when guessing the race and ethnicity of minorities within a block. The proportional guesser would perform better with minorities at the expense of overall accuracy, and so both baselines deserve consideration. 

\subsubsection{Possible improvements to inference accuracy} Two additional points are worth making: first, the respondent's name may also be used to infer race and ethnicity, which we have not addressed in this paper; second, we did not use additional published tables which could be used to reconstruct other features of the 2010 Census HDF.

The feature set $\lbrace race$, $ethnicity \rbrace$ can also be inferred using the name of the target individual, in which case, the name needs to be part of the observed world and counterfactual inferences \citep{fiscella:fremont:2006}. However, since the reconstruction and reidentification algorithms used in this paper do not model this interaction, it is reasonable to omit it from the baseline. What is important for the comparison between the reconstruction and the statistical baseline results is: (1) all else being equal, the estimated privacy cost of being included in the census publications is the difference in inference accuracy (precision) of the reconstruction versus the statistical baseline,\footnote{A large difference between the reconstruction accuracy and the statistical baseline accuracy when less information is used, that is, when name is not used for the race and ethnicity inference, is particularly concerning because it directly indicates that confidentiality has already been breached and that additional information from the census response could be precisely revealed in a larger attack based on all SF1 tables.} and (2) the contrast between the reconstruction and the statistical baseline inferences gives a lower bound on this privacy cost---the actual confidentiality breach could be larger but not smaller.

Second, had the reconstruction used additional variables in the published tables including household composition, the reconstructed data could have included additional attributes that are much harder to predict than race and ethnicity using only statistical information. In those cases, the gap between inference due to reconstruction-abetted reidentification and statistical inference would be much larger. This is another sense in which the experiments we performed likely understate the true privacy cost. Some examples are provided in the supplemental text Section 2.2.

\subsubsection{Methodology for evaluating the statistical baselines} In order to assess the relative accuracy of reidentifications using either the MDG or PRG, we generate two databases, one containing the modal value of the $race \times ethnicity$ distribution in the census block (used by MDG) and the other containing one guess per record using the probabilities proportional to the distribution of $race \times ethnicity$ in the census block (used by PRG). Specifically, we use the HDF to create a frame of $\lbrace$\textit{block, sex, agebin}$\rbrace$ to which we attach the statistical baseline guesses of $\lbrace$\textit{race, ethnicity}$\rbrace$ as follows. For the MDG database, we then compute, for each \textit{block}, the modal value of the $race \times ethnicity$ distribution and attach it to each record in the block.\footnote{Given a tie in the block-level modal $\lbrace$\textit{race, ethnicity}$\rbrace$ or a block population of 1, we attempt to assign the block-group-level modal value. (In the Census Bureau geographic hierarchy, the level between census tract and census block is the block group.) In the rare event that no block group mode can be assigned, either because of a block-group-level tie or a block group population of 1, we assign the block the national modal $\lbrace$\textit{race, ethnicity}$\rbrace$, which is non-Hispanic, White alone. Using this methodology, 95.58\% of blocks (housing 99.44\% of total population) had a unique modal value for $race \times ethnicity$ and 4.42\% of blocks were assigned the block group-level modal value. 0.00\% of blocks required resolution using the national modal value. A similar exercise could be performed using published tables. One could use the tabulations in tables P12 and P14 from SF1 to create a frame of microdata records containing $\lbrace$\textit{block, sex, agebin}$\rbrace$, then use SF1 Tables P8 and P9 to compute the block-level MDG modal values.} For the PRG database, we randomly select a $\lbrace$\textit{race, ethnicity}$\rbrace$ pair for each record, guessing each $\lbrace$\textit{race, ethnicity}$\rbrace$ in proportion to its relative frequency within the block-level $race \times ethnicity$ distribution.\footnote{Similarly to MDG, the public tables can can be used to  proportionally select $\lbrace$\textit{race, ethnicity}$\rbrace$ PRG values.}

We substitute each of these statistical baselines for the reconstructed HDF in the reidentification experiments to generate the baseline statistics. Notice that the \rHDFB, \rHDFBT, MDG, and PRG databases have identical putative match rates using the binned age schema by construction, since all rely on an identical frame of $\lbrace$\textit{block, sex, agebin}$\rbrace$ and both reconstructions perfectly replicate this frame because tables P12 and P14 are fully saturated for $\lbrace$\textit{block, sex, agebin}$\rbrace$.

\subsubsection{Reidentification metrics} Finally, we define several reidentification metrics that capture the accuracy of inferences an attacker can make about the target sensitive characteristics. Putative reidentifications are the records for which the attacker attempts to infer $\lbrace$\textit{race, ethnicity}$\rbrace$. Confirmed reidentifications are the records for which the attacker is correct. The attacker's precision (their potential confidence in the successful inference of $\lbrace$\textit{race, ethnicity}$\rbrace$)  is the ratio of the number of correct inferences to the number of attempted inferences. Thus, we define the putative and confirmed reidentification rates as well as the precision rate as follows:
\begin{equation}
    \label{eq:putative_rate}
    \text{Putative Reidentification Rate} = 100 \times
    \frac{\text{count of putative reidentification records}}{\text{count of attacker records}},
\end{equation}
\begin{equation}
    \label{eq:confirmed_rate}
    \text{Confirmed Reidentification Rate} = 100 \times
    \frac{\text{count of confirmed reidentification records}}{\text{count of attacker records}},    
\end{equation}
\begin{equation}
    \label{eq:precision_rate}
    \text{Reidentification Precision Rate} = 100 \times
    \frac{\text{count of confirmed reidentification records}}{{\text{count of putative reidentification records}}}.
\end{equation}
\noindent We use these statistics to compare our reidentification results with nonstatistical baselines like the HDF itself (i.e., how much better would the reidentification be if the reconstruction were a perfect match to HDF) and with the statistical baselines MDG and PRG. 

In all cases, the attacker model is the same. The attacker begins with tabular summaries produced from a confidential source, the HDF. Then, the attacker reconstructs record-level images of the persons represented in the tables using either the schema \rHDFBT or \rHDFB. Next, the attacker uses a record-level image that contains identifiable names ($pik$, in this work) and linking variables using the same schema as the reconstructed data to perform a record-linkage attack that associates the features \textit{race} and \textit{ethnicity} with each record in the attacker database. Finally, we assess the accuracy of the attack by linking the putative files back to the original confidential data using the features \textit{pik} and \textit{block} for linkage, but assessing the accuracy by comparing the complete set of attributes in the schema.

In summary, we assert that the principal indicator of confidentiality-eroding inference is that the precision rate for inferences based on the reconstructed microdata substantially exceeds the precision rate for inferences based on the statistical guessing baselines. The gain in precision, which we also call the gain in inference accuracy, from the reconstructed data compared to the statistical baseline need not be infinite to constitute a confidentiality breach. The upper-bound on the precision rate of any inference is the precision of that inference based on the CEF itself. We take it as self-evident that releasing every person record in the CEF for the feature set \{\textit{block, sex, age, race, ethnicity}\} is a breach of confidentiality. Additional details can be found in the supplemental text Section 5.3.

\subsection{National results}
\label{subs:reid_natblksz}

Table \ref{tab:reid_national} presents the statistics for putative reidentifications (numerator in Equation \ref{eq:putative_rate}), confirmed reidentifications (numerator in Equation \ref{eq:confirmed_rate}), and precision rates (Equation \ref{eq:precision_rate}) for all data-defined persons in the 2010 Census based on all data-defined persons in either COMRCL or $\text{CEF}_{atkr}$. Detailed results by census block size appear in Table 12 of the supplemental text, the top panel of which is identical to Table \ref{tab:reid_national}.  

The results in Table \ref{tab:reid_national} are refinements of those first released in 2019 \citep{abowd:2019:AAAS}. They show that when the attacker uses the lower-quality COMRCL data, the reconstruction \rHDFBT produces 166,100,000 putative reidentifications (57.9\% putative rate for the data-defined population) of which 68,480,000 are confirmed (23.9\% confirmation rate for the data-defined population) with a reidentification precision rate of 41.2\%. When the attacker uses the high-quality quasi-identifier data in $\text{CEF}_{atkr}$, the reconstruction \rHDFBT yields 267,800,000 putative reidentifications (97.0\% of the data-defined population) of which 208,500,000 are confirmed (75.5\% of the data-defined population) with a precision rate of 77.9\%. 

\begingroup
\footnotesize
\centering
\begin{longtable}{lr|rrrr|rrrr}
    \caption{All Data-Defined Persons: Putative and Confirmed Reidentifications} \\
    \label{tab:reid_national}
     & & \multicolumn{4}{c|}{Attacker ($R$ in Alg. 3): COMRCL} &   \multicolumn{4}{c}{Attacker ($R$ in Alg. 3): $\text{CEF}_{atkr}$} \\
    Data  & Census & Popu-  & Puta- & Con-   & Preci-  & Popu-  & Puta- & Con- & Preci- \\  
    ($L$ in  & Block  & lation & tive  & firmed & sion         & lation & tive  & firmed & sion       \\  
    Alg. 3)  & Size  & ($\times 10^{3}$) & ($\times 10^{3}$)  & ($\times 10^{3}$) & (\%)         & ($\times 10^{3}$) & ($\times 10^{3}$)  & ($\times 10^{3}$) & (\%)      \\ 
    \midrule
    \endfirsthead
    \multicolumn{8}{l}{Table \ref{tab:reid_national} Continued} \\
    \midrule
     & & \multicolumn{4}{c|}{Attacker ($R$ in Alg. 3): COMRCL} &   \multicolumn{4}{c}{Attacker ($R$ in Alg. 3): $\text{CEF}_{atkr}$} \\
    Data  & Census & Popu-  & Puta- & Con-   & Preci-  & Popu-  & Puta- & Con- & Preci- \\  
    ($L$ in  & Block  & lation & tive  & firmed & sion         & lation & tive  & firmed & sion       \\  
    Alg. 3)  & Size  & ($\times 10^{3}$) & ($\times 10^{3}$)  & ($\times 10^{3}$) & (\%)         & ($\times 10^{3}$) & ($\times 10^{3}$)  & ($\times 10^{3}$) & (\%)      \\
    \midrule
    \endhead
    \endfoot
    \multicolumn{10}{l}{Notes: Census Block Size is the population range in the census block. Counts rounded to four}\\
    \multicolumn{10}{l}{significant digits to conform to disclosure limitation requirements. Population for attacker }\\
    \multicolumn{10}{l}{COMRCL is the total number of data-defined records in COMRCL that are also in the CEF }\\
    \multicolumn{10}{l}{universe (see Table \ref{tab:cef_comrcl_match}). Population for attacker $\text{CEF}_{atkr}$ is the total number of data-}\\
    \multicolumn{10}{l}{defined CEF records. This table is identical to supplemental text Table 9.}\\
    \endlastfoot
CEF&All&286,700&167,500&82,760&49.4&276,000&276,000&237,500&86.1 \\ 
HDF&All&286,700&166,100&80,540&48.5&276,000&267,800&228,400&85.3 \\ 
\rHDFBT&All&286,700&166,100&68,480&41.2&276,000&267,800&208,500&77.9 \\ 
\rHDFB&All&286,700&166,100&67,450&40.6&276,000&267,800&203,100&75.9 \\ 
MDG&All&286,700&166,100&76,270&45.9&276,000&267,800&205,100&76.6 \\ 
PRG&All&286,700&166,100&66,260&39.9&276,000&267,800&177,700&66.3 \\ 
\midrule
\end{longtable}
\endgroup

The Census Bureau released similar reidentification results as part of litigation surrounding the 2020 Census and to its Scientific Advisory Committee between 2019 and 2022 \citep{abowd2_fairlines_appendixB, hawes:2022, abowd:hawes:2023} based on earlier versions of the models presented in this paper. Some researchers noted that baselines similar to MDG and PRG could produce reidentifications apparently comparable to those of \rHDFBT \citep{ruggles:vanriper:2022,francis:2022}. The rows MDG and PRG in the COMRCL panel confirm that claim but only, as we shall see, if one ignores nonmodal and vulnerable populations.  By construction, these two strategies have the same overall putative reidentifications as \rHDFBT, and their confirmation and precision rates are also comparable. Comparison to statistical baselines at the national level highlights the fact that national reidentification rates, while informative about the scale of the match (e.g., 68,480,000 individuals had confirmed \{\textit{race, ethnicity}\} in the COMRCL experiment using \rHDFBT), are difficult to interpret in terms of confidentiality violation. The misleading similarity between statistical baselines and \rHDFBT requires using leave-one-out reasoning to resolve, which we do below. \rHDFBT\ is able to reidentify the \textit{race} and \textit{ethnicity} of nonmodal individuals that are unique in the linking quasi-identifiers \{\textit{block, sex, agebin}\} in blocks with  zero solution variability, where MDG and PRG completely fail---a clear demonstration of confidentiality violation.\footnote{Since this paper is primarily a proof of concept, a larger-scale attack using more tables would result in less solution variability and potentially more population uniques.}

Surprisingly, releasing the binned-age schema from the CEF itself would produce only relatively modest increase in the putative reidentification rates compared to HDF, \rHDFBT, \rHDFB or either baseline, as shown in the first row of Table \ref{tab:reid_national} for both COMRCL and $\text{CEF}_{atkr}$. As noted in Table \ref{tab:cef_comrcl_match}, COMRCL data did not align well with the data collected in the 2010 Census. But even if the attacker used high-quality input data, as in $\text{CEF}_{atkr}$, releasing the CEF in this schema would still produce only modestly more putative reidentifications, and releasing the HDF  would produce exactly the same putative reidentifications as \rHDFBT, \rHDFB, or either baseline. From the viewpoint of records at risk for record-linkage reidentification, the reconstructed microdata and the statistical baselines accurately replicate HDF. The confirmation and precision rates for CEF and HDF are marginally higher than those for \rHDFBT, \rHDFB and MDG, and substantially better than PRG.

\subsection{Detailed results for nonmodal race-ethnicity records by census block size}\label{subs:reid_nonmdl}

\begin{figure}[ht]
    \centering
    \includegraphics[width=\textwidth]{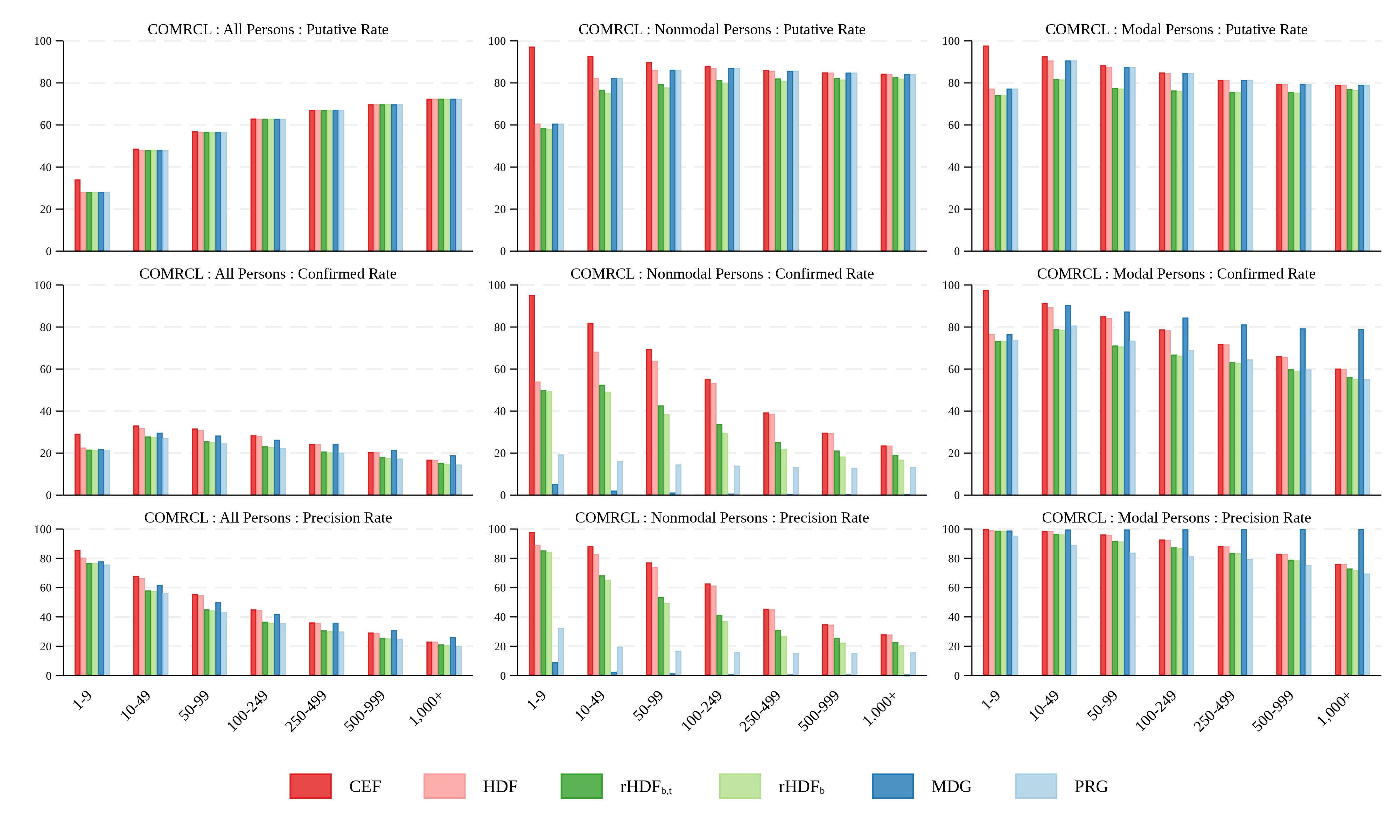}
    \\
    \caption{Comparison of Putative Reidentification Rates, Confirmed Reidentification Rates, and Reidentification Precision Rates for COMRCL by Census Block Size  \\Notes: The denominator used in the first column is the COMRCL Population column in supplemental text Table 12, which totals to 286,700,000. The denominator in the second column is the COMRCL population in supplemental text Table 13, and the denominator in the third column is the COMRCL population in supplemental text Table 14. The denominators in the second and third columns sum to 106,300,000 because only that subset of COMRCL records can be classified as modal and nonmodal from the CEF. See Table \ref{tab:cef_comrcl_match}.  This figure is identical to Figure 3 in the supplemental text.}
    \label{fig:COMRCL}
\end{figure}

\begin{figure}[ht]
    \centering
    \includegraphics[width=\textwidth]{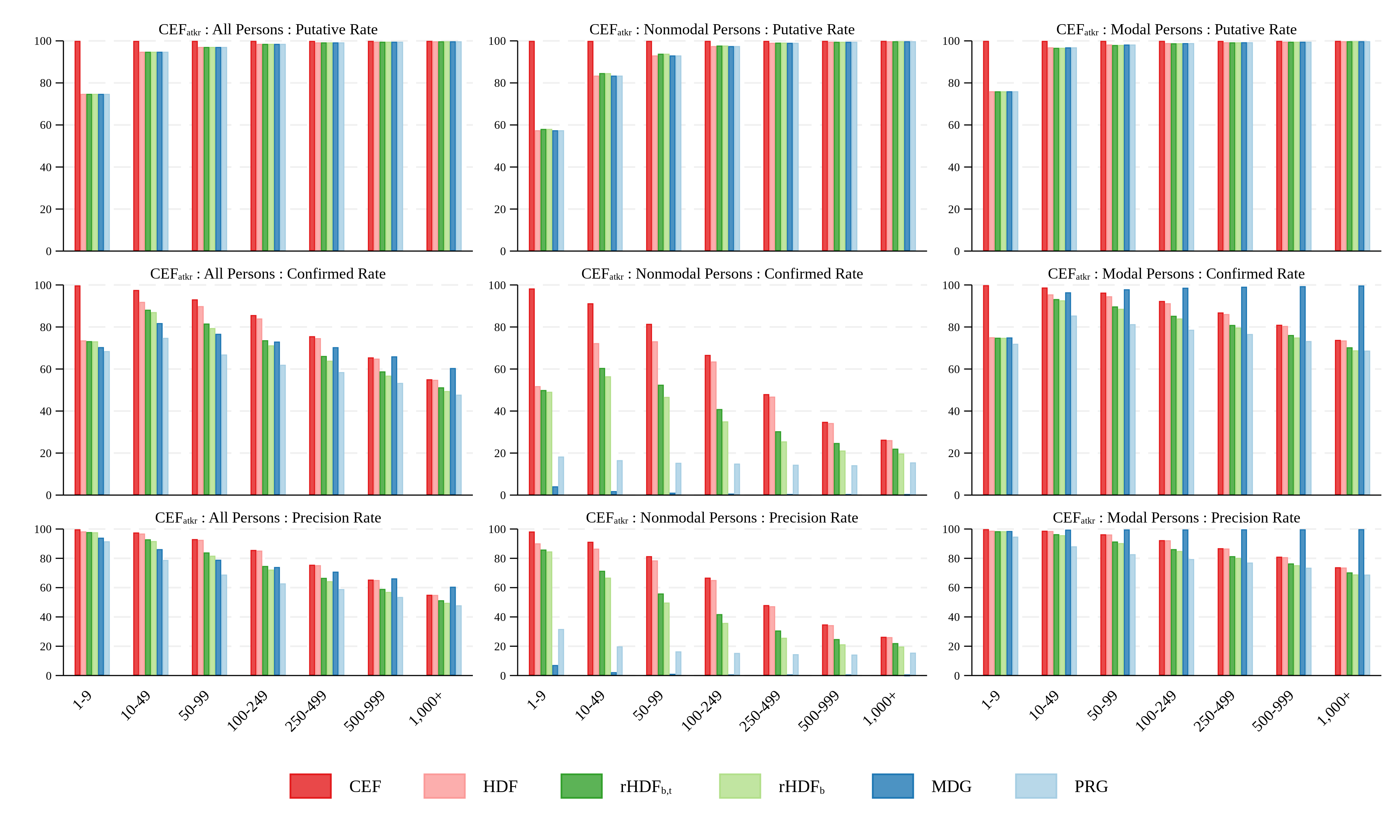}
    \\
    \caption{Comparison of Putative Reidentification Rates, Confirmed Reidentification Rates, and Reidentification Precision Rates for $\text{CEF}_{atkr}$ by Census Block Size \\Notes: The denominator used in the first column is the $\text{CEF}_{atkr}$ Population column in supplemental text Table 12, which totals to 276,000,000. The denominator in the second column is the $\text{CEF}_{atkr}$ population in supplemental text Table 13, and the denominator in the third column is the $\text{CEF}_{atkr}$ population in supplemental text Table 14. The denominators in the second and third columns sum to 276,000,000 because all records in $\text{CEF}_{atkr}$ can be classified as modal and nonmodal from the CEF. See Table \ref{tab:cef_comrcl_match}. This figure is identical to Figure 4 in the supplemental text.}
    \label{fig:CEFatkr}
\end{figure}

To assess the impact of confidentiality-violating inference from reconstruction against the generalizable inferences made by the statistical guessers, we compare the reidentification results of the different models for the modal and nonmodal subsets of the population in census blocks of varying population size. Figures \ref{fig:COMRCL} and \ref{fig:CEFatkr} summarize these reidentification results, by census block size, for the lower-quality COMRCL and higher-quality $\text{CEF}_{atkr}$ data, respectively. The detailed results are shown in supplemental text Tables 12, 13, and 14. The first column of both figures shows the results for all data-defined persons by census block size. As summarized in the national results discussion, except for the lowest population blocks (1 to 9 persons), there are very few discernible differences between the CEF, HDF, \rHDFBT, \rHDFB, MDG, and PRG results.

Comparing the results for nonmodal data-defined persons (middle column) with those of modal data-defined persons (last column) reveals the power in the reconstructed microdata vis-\`{a}-vis the statistical baselines. For comparing and contrasting the nonmodal and modal persons results, the interpretation does not depend on the attacker database (see Section 7.2 of the supplemental text for the only substantive analytic difference between these attacker files), so we characterize them generically. Because the statistical guessing models MDG and PRG use the known racial and ethnic characteristics of the block populations to guess (via statistical inference) the race and ethnicity for each record, it is unsurprising that both models perform well for the modal populations of those blocks. In these figures, it is the nonmodal data-defined persons that we consider to be the vulnerable population, as their race and ethnicity could only be accurately revealed through confidentiality-violating inference. 
 
Except for blocks with populations of 1-9 persons, CEF, HDF, \rHDFBT, \rHDFB, MDG, and PRG putatively reidentify nonmodal persons at essentially the same rates because SF1 Tables P12 and P14 contain all the information used for putative reidentifications. It is the confirmation and precision rates that reveal the contrast with the modal person subpopulation. In blocks with 1-9 persons, the HDF, \rHDFBT, and \rHDFB have essentially identical confirmation and precision rates, whereas the MDG baseline is nearly zero for both confirmation and precision rates, and the PRG baseline is greater than the MDG but substantially less than either \rHDFBT or \rHDFB. As the population in the census block increases, the confirmation and precision rates of \rHDFBT and \rHDFB decline but remain high, much closer to the rates for CEF and HDF than MDG (always essentially zero) and approaching proportional guessing only for the most populous census blocks.  Even though there are 125 possible nonmodal \{\textit{race, ethnicity}\} combinations, the reidentification precision rate of the reconstructed data is greater than 50\% for 23,765,000 persons living in census blocks with populations of 1-99 persons (more than $1/3$ of all nonmodal data-defined persons) compared to essentially zero for the modal guesser. For these 23,765,000 persons, the smallest gain in the reidentification precision rate relative to the proportional guesser is 32.6 percentage points.\footnote{Supplemental text Table 13, Census Block Size 50-99, COMRCL attacker, \rHDFB reconstruction.} This enormous gain in accuracy occurs only because the nonmodal person's data were used in the SF1 tabulations. 

When the vulnerable population is defined as persons who differ from the predominant characteristics in small neighborhoods, in this case nonmodal \{\textit{race, ethnicity}\} persons in lower-population census blocks, the statistical baselines either fail completely (MDG) or fare no better than chance (PRG), but the reconstructed microdata are correct at rates that approach the correctness of the confidential data themselves as the neighborhood population decreases.

Additional details for the nonmodal vulnerable persons can be found in supplemental text Section 7.2.

\subsection{Results for the most vulnerable populations}
\label{subs:vulnerable_populations}

We now discuss the results for the most tightly constrained definition of the vulnerable population---nonmodal data-defined persons living in blocks with  zero solution variability whose reconstructed records are population uniques for \{\textit{block, sex, agebin}\}.  Panel A of Table \ref{tab:nm_solvar} shows the results for nonmodal persons living in  zero solution variability blocks and Panel B shows the results for the subset of these persons who are population unique on \{\textit{block, sex, agebin}\}. Recalling that population uniqueness on \{\textit{block, sex, agebin}\} implies population uniquness on \{\textit{block, sex, age}\} and that  zero solution variability implies that the record in the reconstructed data is provably the same as the record in HDF without referencing any confidential data, Table \ref{tab:nm_solvar} Panel B establishes that the reidentification precision rates of \rHDFBT and HDF are essentially identical: 95.4 versus 96.4 in COMRCL and 95.2 versus 95.6 in $\text{CEF}_{atkr}$. These reidentification precisions are far in excess of the modal guesser's (3.3 for COMRCL and 2.5 for $\text{CEF}_{atkr}$) or the proportional guesser's (20.9 and 20.2, resp.). Thus, the correct \{\textit{race, ethnicity}\} has been exposed for more than 19 of 20 persons in the population of nonmodal, 0-$solvar$, \{\textit{block, sex, agebin}\} uniques---essentially the same exposure as publishing the HDF itself, and only five percentage points worse than publishing the CEF itself. 

Statistical baselines are either almost always wrong (for the modal guesser) or wrong for 4 of 5 persons (for the proportional guesser) in this vulnerable population. These results establish that the published SF1 tables enable clear confidentiality breaches. The high statistical precision of the reconstructed microdata is only possible because the target person's data were used in the SF1 tabulations. Had the target person not participated in the 2010 Census, these inferences would be impossible. Therefore, the results in Table \ref{tab:nm_solvar} definitively establish that SF1 permits very high accuracy reidentifications of population-unique individuals with characteristics that differ from their neighbors. These are not statistical inferences, because they disappear if that person does not participate in the census. They are confidentiality breaches. 

Additional details and discussion of Table \ref{tab:nm_solvar} can be found in the supplemental text Section 8.

\begingroup
\footnotesize
\centering
\begin{longtable}{l|rrrr|rrrr}
    \caption{Putative Reidentifications, Confirmed Reidentifications, and Precision Rates for Nonmodal Persons in Blocks with Zero Solution Variability} \\
    \label{tab:nm_solvar}
    &  \multicolumn{4}{c|}{Attacker ($R$ in Alg. 3): COMRCL} &   \multicolumn{4}{c}{Attacker ($R$ in Alg. 3): $\text{CEF}_{atkr}$} \\
    & Popu-       & Puta-         & Con-   & Preci-      & Popu- & Puta-         & Con-          & Preci-\\  
    Data ($L$ in & lation       & tive         & firmed    & sion      & lation & tive         & firmed          & sion\\  
    Algorithm 3)  & ($\times 10^{3}$) & ($\times 10^{3}$) & ($\times 10^{3}$) & (\%)  & ($\times 10^{3}$)      & ($\times 10^{3}$) & ($\times 10^{3}$) & (\%) \\
    \midrule
    \endfirsthead
    \multicolumn{8}{l}{Table \ref{tab:nm_solvar} Continued} \\
    \midrule
    &  \multicolumn{4}{c|}{Attacker ($R$ in Alg. 3): COMRCL} &   \multicolumn{4}{c}{Attacker ($R$ in Alg. 3): $\text{CEF}_{atkr}$} \\
    & Popu-       & Puta-         & Con-   & Preci-      & Popu- & Puta-         & Con-          & Preci-\\  
    Data ($L$ in & lation       & tive         & firmed    & sion      & lation & tive         & firmed          & sion\\  
    Algorithm 3)  & ($\times 10^{3}$) & ($\times 10^{3}$) & ($\times 10^{3}$) & (\%)  & ($\times 10^{3}$)      & ($\times 10^{3}$) & ($\times 10^{3}$) & (\%) \\
    \midrule
    \endhead
    \endfoot
    \multicolumn{9}{l}{Notes: Counts rounded to four significant digits to conform to disclosure limitation}\\
    \multicolumn{9}{l}{ requirements. The rows labeled $\text{rMDF}_{b,t}$ and MDF (light gray highlight) are the analogs}\\
    \multicolumn{9}{l}{of \rHDFBT and HDF, respectively, when the 2020 Census DAS is applied to the 2010 CEF. }\\
    \multicolumn{9}{l}{The rows labeled $\text{rSWAPLo}_{b,t}$ and $\text{rSWAPHi}_{b,t}$ (medium gray highlight) are the analogs }\\
    \multicolumn{9}{l}{of \rHDFBT when alternative swapping parameters are used. Population for attacker}\\
    \multicolumn{9}{l}{COMRCL is data-defined nonmodal person records in COMRCL that match CEF}\\
    \multicolumn{9}{l}{ on the feature set \{\textit{pik, block}\}. Population for attacker $\text{CEF}_{atkr}$ is all data-defined}\\
    \multicolumn{9}{l}{  nonmodal persons in CEF. This table is identical to supplemental text Table 10. }\\
    \endlastfoot
  & \multicolumn{8}{c}{Panel A: Nonmodal Persons} \\
CEF&2,098&1,925&1,633&84.8&6,517&6,517&5,727&87.9 \\ 
HDF&2,098&1,646&1,278&77.7&6,517&5,189&4,209&81.1 \\ 
\rHDFBT&2,098&1,557&1,009&64.8&6,517&5,305&3,634&68.5 \\ 
\rHDFB&2,098&1,516&914&60.3&6,517&5,304&3,369&63.5 \\ 
MDG&2,098&1,646&33&2.0&6,517&5,189&88&1.7 \\ 
PRG&2,098&1,646&274&16.6&6,517&5,189&890&17.2 \\ 
\rowcolor{Gray} MDF&2,098&593&103&17.3&6,517&1,866&353&18.9 \\ 
\rowcolor{Gray} $\text{rMDF}_{b,t}$&2,098&593&103&17.3&6,517&1,866&352&18.9 \\ 
\rowcolor{Medgray} $\text{rSWAPLo}_{b,t}$&2,098&1,733&1,236&71.3&6,517&6,242&4,693&75.2 \\ 
\rowcolor{Medgray} $\text{rSWAPHi}_{b,t}$&2,098&1,255&680&54.1&6,517&4,270&2,482&58.1 \\ 
&  \multicolumn{8}{c}{Panel B: Nonmodal Uniques on $\{block,sex,agebin\}$} \\
CEF&908&834&834&100.0&3,364&3,364&3,364&100.0 \\ 
HDF&908&649&625&96.4&3,364&2,418&2,311&95.6 \\ 
\rHDFBT&908&587&560&95.4&3,364&2,418&2,301&95.2 \\ 
\rHDFB&908&565&537&95.0&3,364&2,418&2,237&92.5 \\ 
MDG&908&649&21&3.3&3,364&2,418&61&2.5 \\ 
PRG&908&649&136&20.9&3,364&2,418&488&20.2 \\ 
\rowcolor{Gray} MDF&908&193&42&21.6&3,364&671&147&21.8 \\ 
\rowcolor{Gray} $\text{rMDF}_{b,t}$&908&193&42&21.7&3,364&671&147&21.8 \\ 
\rowcolor{Medgray} $\text{rSWAPLo}_{b,t}$&908&723&710&98.2&3,364&3,194&3,111&97.4 \\ 
\rowcolor{Medgray} $\text{rSWAPHi}_{b,t}$&908&485&371&76.4&3,364&2,014&1,568&77.9 \\ 
\midrule
\end{longtable}
\endgroup

\section{Did the 2020 Disclosure Avoidance System address the\\ disclosure risk better than feasible alternatives?}
\label{sec:2020_das}

Our final discussion addresses whether any SDL method can effectively defend against reiden\-ti\-fi\-ca\-tion attacks while simultaneously allowing publication of data that are fit for use for their intended purposes. \citet{christ:radway:bellovin:2020} demonstrated that the Census Bureau's differentially private disclosure avoidance system could provide fitness-for-use in the redistricting application comparable to swapping but with superior confidentiality protection. They also noted that ``[s]wapping places a disproportionate privacy burden on minority groups,'' while differential privacy protections apply to all sub-populations. These results, which we confirm in this paper, are in contrast to the erroneous conclusions drawn by \citet{kenny:et:al:2021} who failed to distinguish between generalizable and privacy-violating inferences \citep[Supplement p. 8]{jarmin:etal:2023}.  \citet{wright:irimata:2021} demonstrated that the 2020 Census Redistricting Data (P.L. 94-171) Summary File was fit for use in redistricting and scrutiny under Section 2 of the 1965 Voting Rights Act \citep{vra}.

On October 20, 2022, DSEP approved the final production settings for the 2020 Census Disclosure Avoidance System (2020 DAS) as applied to the Demographic and Housing Characteristics (DHC) File, the successor to the 2010 SF1 used in this paper. DHC was produced using the algorithms described in \citet{Abowd20222020} (for the redistricting data tables) and \citet{cumings-menon:etal:2023} (for the remainder of the tables in the DHC). 

The statistical summaries presented for the DSEP decision included an analysis of the effectiveness of the same reconstruction-abetted reidentification attack studied here based on the 2010 Census. The analysis compared the confidentiality protections from the original SF1 tables shown in Table \ref{tab:SF1_tables} and Figures \ref{fig:COMRCL} and \ref{fig:CEFatkr} with the protection afforded by the 2020 DAS, when executed with the 2020 Census DHC production settings applied to the 2010 Census data, and the protection afforded by housing unit-level data swapping at rates of 5\% and $50\%$, i.e., swapping 5\% or $50\%$ of all housing units to a new location in a different block, tract, or elsewhere in the state. Additional details can be found in the supplemental text Appendix C.

Table \ref{tab:agree_blockpop_mdf} shows the agreement rates overall and by census block size between the 2010 CEF and
\begin{itemize}
    \item HDF,
    \item \rHDFBT,
    \item \rHDFB,
    \item $\text{rMDF}_{b,t}$---reconstructed Microdata Detail File (MDF), \citep{Abowd20222020,cumings-menon:etal:2023}, the production version output of the 2020 DAS for the DHC when applied to 2010 Census data and processed into the equivalents of the Table \ref{tab:SF1_tables} statistics,
    \item MDF---actual MDF from the same 2020 DAS output; i.e., not processed into tables, so there is no reconstruction of the microdata---these are the actual microdata from which DHC was tabulated,
    \item $\text{rSWAPLo}_{b,t}$---the results of the 5\% swap-rate experiment when processed through our reconstruction-abetted reidentification attack,
    \item $\text{rSWAPHi}_{b,t}$---the results of the 50\% swap-rate experiment when similarly processed.
\end{itemize}
The $\text{rMDF}_{b,t}$, MDF, and  $\text{rSWAPHi}_{b,t}$ reduce the agreement with the 2010 CEF substantially, with the $\text{rMDF}_{b,t}$ and MDF reducing agreement more than $\text{rSWAPHi}_{b,t}$. The $\text{rSWAPLo}_{b,t}$ agrees with the CEF at essentially the same rate as \rHDFBT except for blocks with populations of 1-9 persons, where HDF, \rHDFBT, and \rHDFB all agree with the CEF for 74.0\% of records using binned age while $\text{rSWAPLo}_{b,t}$ agrees with 94.9\% of these records. The experimental swap methodology did not single-out low population blocks for relatively greater swapping.

\begingroup
\footnotesize
\centering
\begin{longtable}{lr|r|rr|rr}
    \caption{Selected Reconstruction Agreement Statistics with Comparisons to Output from the 2020 DAS and Specially Swapped Versions of the CEF Using the 2010 Census as Input by Census Block Size} \\
    \label{tab:agree_blockpop_mdf}
    & Census & & & & & \\
    Data ($L-R$ in & Block & Population & \multicolumn{2}{c|}{Agreement ($\times 10^{3}$)} & \multicolumn{2}{c}{Agreement (\%)}  \\
    Algorithm 2) & Size & ($\times 10^{3}$) & Exact Age & Binned Age & Exact Age & Binned Age \\
    \midrule
    \endfirsthead
    \multicolumn{7}{l}{Table \ref{tab:agree_blockpop_mdf} Continued} \\
    \midrule
    & Census & & & & & \\
    Data ($L-R$ in & Block & Population & \multicolumn{2}{c|}{Agreement ($\times 10^{3}$)} & \multicolumn{2}{c}{Agreement (\%)}  \\
    Algorithm 2) & Size & ($\times 10^{3}$) & Exact Age & Binned Age & Exact Age & Binned Age \\
    \midrule
    \endhead
    \endfoot
    \multicolumn{7}{l}{Notes: Census Block Size is the population range in the census block. Counts rounded to }\\
    \multicolumn{7}{l}{four significant digits, except block populations, to conform to disclosure limitation }\\
    \multicolumn{7}{l}{requirements. Agreement percentages use the block populations in that row as the base. }\\
    \multicolumn{7}{l}{The rows labeled $\text{rMDF}_{b,t}$ and MDF (light gray highlight) are the analogs of \rHDFBT } \\
    \multicolumn{7}{l}{and HDF, resp., when the 2020 Census DAS is applied to the 2010 CEF. Rows labeled }\\
    \multicolumn{7}{l}{$\text{rSWAPLo}_{b,t}$ and $\text{rSWAPHi}_{b,t}$ (medium gray highlight) are the results of applying the full}\\
    \multicolumn{7}{l}{reconstruction-abetted reidentification attack to the specially swapped versions of CEF}\\
    \multicolumn{7}{l}{described in the supplemental text Section 9 and Appendix C. This table is identical to}\\
    \multicolumn{7}{l}{Table 11 in the supplemental text.}\\
    \endlastfoot
HDF-CEF&All&308,746&297,200&297,600&96.3&96.4 \\ 
\rHDFBT-CEF&All&308,746&143,800&283,600&46.6&91.9 \\ 
\rHDFB-CEF&All&308,746&132,200&276,900&42.8&89.7 \\ 
\rowcolor{Gray} $\text{rMDF}_{b,t}$-CEF&All&308,746&58,520&113,100&18.9&36.6 \\ 
\rowcolor{Gray} MDF-CEF&All&308,746&75,950&113,300&24.6&36.7 \\ 
\rowcolor{Medgray} $\text{rSWAPLo}_{b,t}$ (5\%)-CEF&All&308,746&144,500&281,200&46.8&91.1 \\ 
\rowcolor{Medgray} $\text{rSWAPHi}_{b,t}$ (50\%)-CEF&All&308,746&100,800&193,200&32.7&62.6 \\ 
HDF-CEF&1-9&8,070&5,866&5,973&72.7&74.0 \\ 
\rHDFBT-CEF&1-9&8,070&2,419&5,971&30.0&74.0 \\ 
\rHDFB-CEF&1-9&8,070&2,325&5,968&28.8&74.0 \\ 
\rowcolor{Gray} $\text{rMDF}_{b,t}$-CEF&1-9&8,070&232&647&2.9&8.0 \\ 
\rowcolor{Gray} MDF-CEF&1-9&8,070&276&647&3.4&8.0 \\ 
\rowcolor{Medgray} $\text{rSWAPLo}_{b,t}$ (5\%)-CEF&1-9&8,070&3,278&7,660&40.6&94.9 \\ 
\rowcolor{Medgray} $\text{rSWAPHi}_{b,t}$ (50\%)-CEF&1-9&8,070&1,803&4,280&22.3&53.0 \\ 
HDF-CEF&10-49&67,598&63,460&63,580&93.9&94.1 \\ 
\rHDFBT-CEF&10-49&67,598&29,500&62,870&43.6&93.0 \\ 
\rHDFB-CEF&10-49&67,598&28,990&62,330&42.9&92.2 \\ 
\rowcolor{Gray} $\text{rMDF}_{b,t}$-CEF&10-49&67,598&4,999&12,330&7.4&18.2 \\ 
\rowcolor{Gray} MDF-CEF&10-49&67,598&6,216&12,320&9.2&18.2 \\ 
\rowcolor{Medgray} $\text{rSWAPLo}_{b,t}$ (5\%)-CEF&10-49&67,598&30,320&63,370&44.9&93.8 \\ 
\rowcolor{Medgray} $\text{rSWAPHi}_{b,t}$ (50\%)-CEF&10-49&67,598&18,110&38,330&26.8&56.7 \\ 
HDF-CEF&50-99&69,073&66,560&66,630&96.4&96.5 \\ 
\rHDFBT-CEF&50-99&69,073&31,280&64,330&45.3&93.1 \\ 
\rHDFB-CEF&50-99&69,073&30,600&63,130&44.3&91.4 \\ 
\rowcolor{Gray} $\text{rMDF}_{b,t}$-CEF&50-99&69,073&8,350&18,830&12.1&27.3 \\ 
\rowcolor{Gray} MDF-CEF&50-99&69,073&10,670&18,820&15.5&27.2 \\ 
\rowcolor{Medgray} $\text{rSWAPLo}_{b,t}$ (5\%)-CEF&50-99&69,073&31,190&63,300&45.2&91.6 \\ 
\rowcolor{Medgray} $\text{rSWAPHi}_{b,t}$ (50\%)-CEF&50-99&69,073&20,200&41,180&29.2&59.6 \\ 
HDF-CEF&100-249&80,021&78,370&78,420&97.9&98.0 \\ 
\rHDFB-CEF&100-249&80,021&34,690&71,940&43.4&89.9 \\ 
\rHDFBT-CEF&100-249&80,021&36,840&73,810&46.0&92.2 \\ 
\rowcolor{Gray} $\text{rMDF}_{b,t}$-CEF&100-249&80,021&15,030&30,810&18.8&38.5 \\ 
\rowcolor{Gray} MDF-CEF&100-249&80,021&19,750&30,790&24.7&38.5 \\ 
\rowcolor{Medgray} $\text{rSWAPLo}_{b,t}$ (5\%)-CEF&100-249&80,021&36,310&71,880&45.4&89.8 \\ 
\rowcolor{Medgray} $\text{rSWAPHi}_{b,t}$ (50\%)-CEF&100-249&80,021&25,740&50,530&32.2&63.2 \\ 
HDF-CEF&250-499&42,911&42,320&42,340&98.6&98.7 \\ 
\rHDFB-CEF&250-499&42,911&18,170&37,960&42.3&88.5 \\ 
\rHDFBT-CEF&250-499&42,911&20,750&39,240&48.3&91.4 \\ 
\rowcolor{Gray} $\text{rMDF}_{b,t}$-CEF&250-499&42,911&12,220&22,570&28.5&52.6 \\ 
\rowcolor{Gray} MDF-CEF&250-499&42,911&16,290&22,600&38.0&52.7 \\ 
\rowcolor{Medgray} $\text{rSWAPLo}_{b,t}$ (5\%)-CEF&250-499&42,911&20,470&38,250&47.7&89.2 \\ 
\rowcolor{Medgray} $\text{rSWAPHi}_{b,t}$ (50\%)-CEF&250-499&42,911&15,830&29,030&36.9&67.7 \\ 
HDF-CEF&500-999&27,029&26,720&26,740&98.9&98.9 \\ 
\rHDFBT-CEF&500-999&27,029&14,220&24,550&52.6&90.8 \\ 
\rHDFB-CEF&500-999&27,029&11,380&23,480&42.1&86.9 \\ 
\rowcolor{Gray} $\text{rMDF}_{b,t}$-CEF&500-999&27,029&10,280&17,210&38.0&63.7 \\ 
\rowcolor{Gray} MDF-CEF&500-999&27,029&13,540&17,310&50.1&64.0 \\ 
\rowcolor{Medgray} $\text{rSWAPLo}_{b,t}$ (5\%)-CEF&500-999&27,029&14,090&24,060&52.1&89.0 \\ 
\rowcolor{Medgray} $\text{rSWAPHi}_{b,t}$ (50\%)-CEF&500-999&27,029&11,480&19,130&42.5&70.8 \\ 
HDF-CEF&1,000+&14,043&13,930&13,940&99.2&99.3 \\ 
\rHDFBT-CEF&1,000+&14,043&8,835&12,870&62.9&91.7 \\ 
\rHDFB-CEF&1,000+&14,043&6,009&12,120&42.8&86.3 \\ 
\rowcolor{Gray} $\text{rMDF}_{b,t}$-CEF&1,000+&14,043&7,407&10,670&52.8&76.0 \\ 
\rowcolor{Gray} MDF-CEF&1,000+&14,043&9,204&10,820&65.5&77.0 \\ 
\rowcolor{Medgray} $\text{rSWAPLo}_{b,t}$ (5\%)-CEF&1,000+&14,043&8,798&12,720&62.7&90.6 \\ 
\rowcolor{Medgray} $\text{rSWAPHi}_{b,t}$ (50\%)-CEF&1,000+&14,043&7,678&10,730&54.7&76.4 \\ 
\midrule
\end{longtable}
\endgroup

\begin{figure}[ht]
    \centering
    \includegraphics[width=\textwidth]{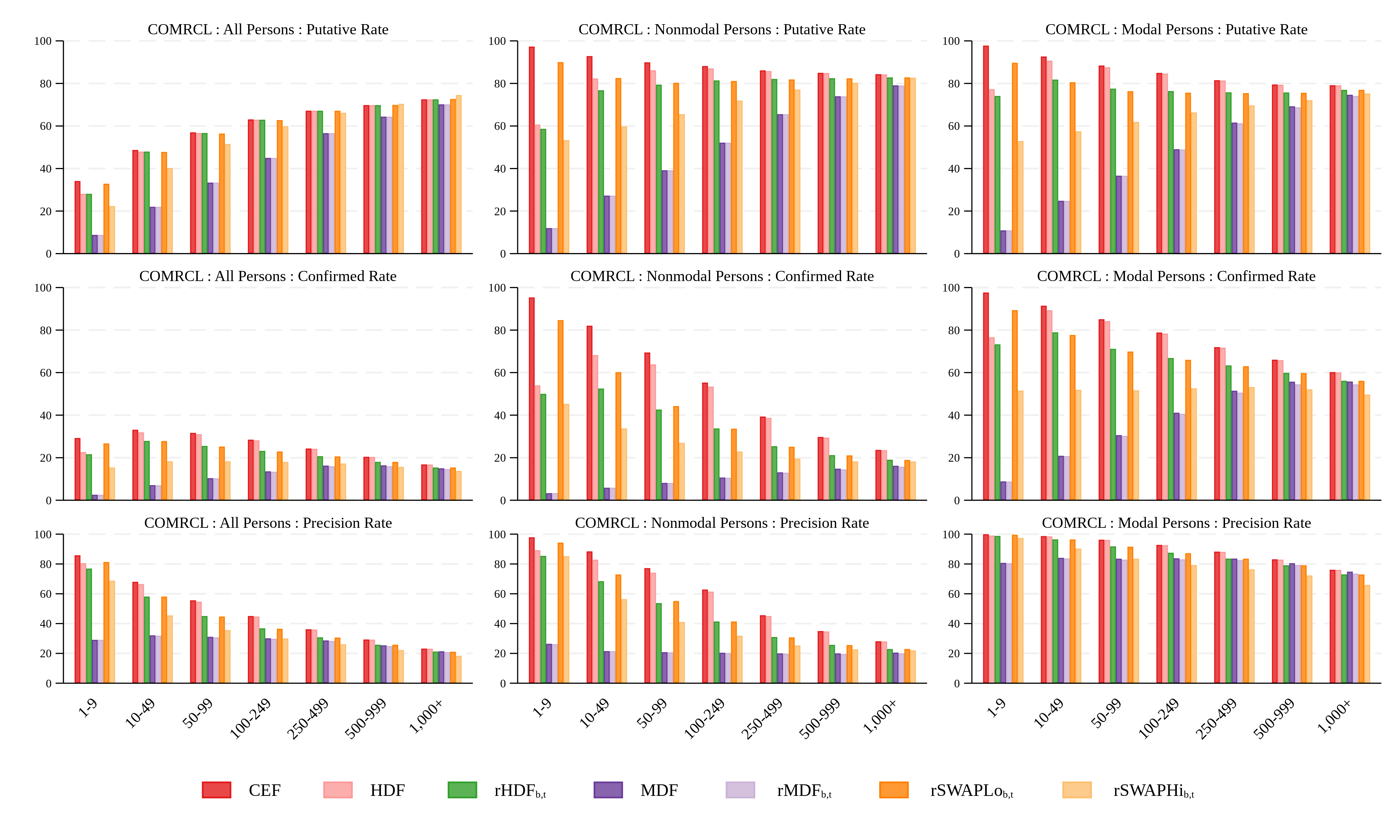}
    \\
    \caption{Comparison of Putative Reidentification Rates, Confirmed Reidentification Rates, and Reidentification Precision Rates for Alternative Disclosure Limitation Implementations Applied to the 2010 Census Edited File for Attacker COMRCL by Census Block Size \\Notes: The denominator used in the first column is the COMRCL Population column in supplemental text Table 15, which totals to 286,700,000. The denominator in the second column is the COMRCL population in supplemental text Table 16, and the denominator in the third column is the COMRCL population in supplemental text Table 17. The denominators in the second and third columns sum to 106,300,000 because only that subset of COMRCL records can be classified as modal and nonmodal from the CEF. See Table \ref{tab:cef_comrcl_match}. This figure is identical to Figure 5 in the supplemental text.}
    \label{fig:COMRCL_mdf}
\end{figure}

\begin{figure}[ht]
    \centering
    \includegraphics[width=\textwidth]{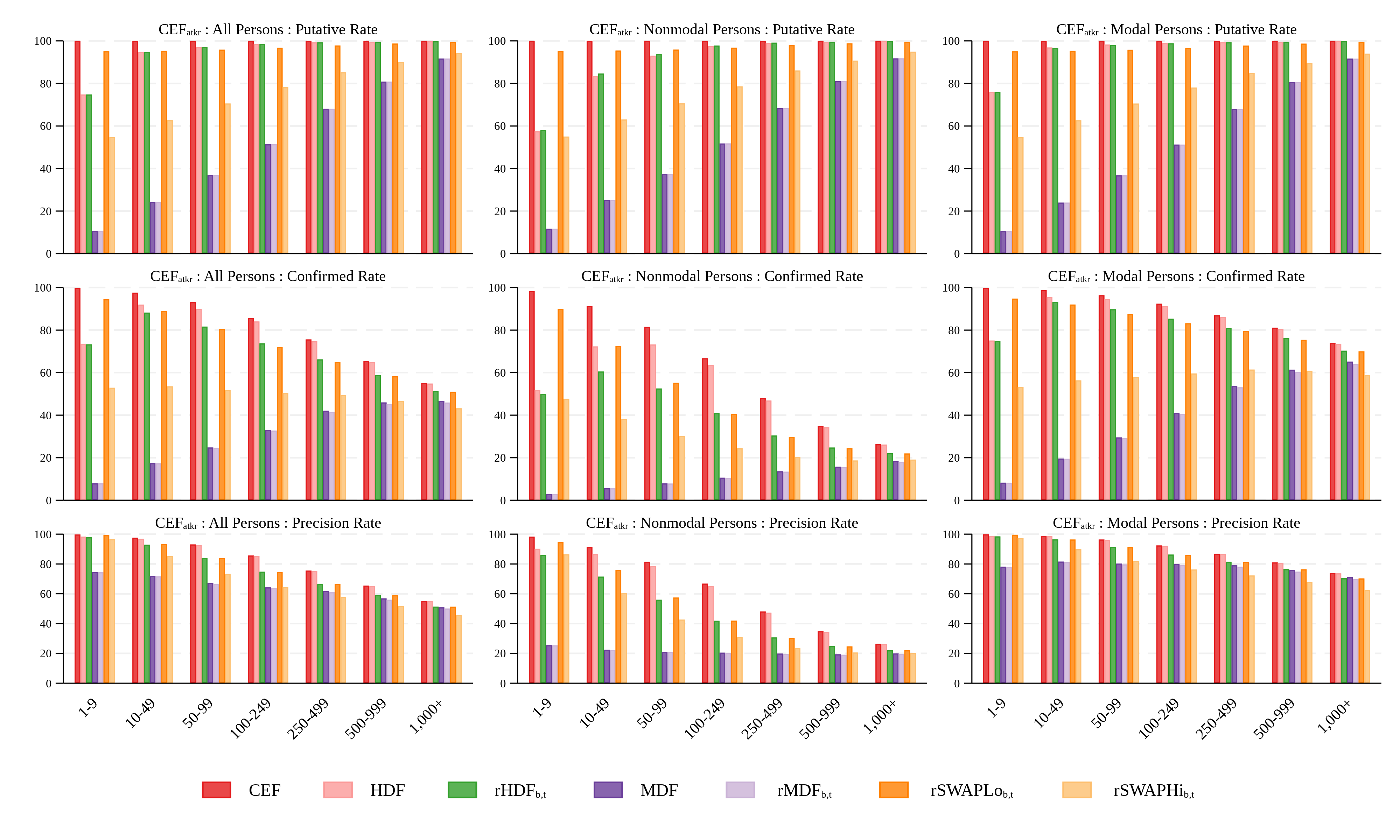}
    \\
    \caption{Comparison of Putative Reidentification Rates, Confirmed Reidentification Rates, and Reidentification Precision Rates for Alternative Disclosure Limitation Implementations Applied to the 2010 Census Edited File for Attacker  $\text{CEF}_{atkr}$ by Census Block Size \\Notes: The denominator used in the first column is the $\text{CEF}_{atkr}$ Population column in supplemental text Table 15, which totals to 276,000,000. The denominator in the second column is the $\text{CEF}_{atkr}$ population in supplemental text Table 16, and the denominator in the third column is the $\text{CEF}_{atkr}$ population in supplemental text Table 17. The denominators in the second and third columns sum to 276,000,000 because all records in $\text{CEF}_{atkr}$ can be classified as modal and nonmodal from the CEF. See Table \ref{tab:cef_comrcl_match}. This figure is identical to Figure 6 in the supplemental text.}
    \label{fig:CEFatkr_mdf}
\end{figure}

Figures \ref{fig:COMRCL_mdf} and \ref{fig:CEFatkr_mdf} compare the putative reidentification, confirmed reidentification, and reidentification precision rates by census block size when the attacker is COMRCL or $\text{CEF}_{atkr}$, respectively. Their interpretation is identical to the interpretation of Figures \ref{fig:COMRCL} and \ref{fig:CEFatkr}. Detailed statistics can be found in the supplemental text Tables 15, 16, and 17.

Only the output of the 2020 DAS ($\text{rMDF}_{b,t}$ and MDF) and $\text{rSWAPHi}_{b,t}$ succeed in reducing putative reidentification rates because only those SDL treatments introduce substantial noise in the quasi-identifiers \{\textit{block, sex, agebin}\}. Without noise in those quasi-identifiers, putative reidentification rates will always be very close to those of the CEF.  
  
The second important feature of Figures \ref{fig:COMRCL_mdf} and \ref{fig:CEFatkr_mdf} is the very low, nearly uniform by block size, reidentification precision rates for $\text{rMDF}_{b,t}$ and MDF for the nonmodal data-defined persons (middle column). These are the only SDL treatments shown in the figures that accomplish this reduction. As the statistics in supplemental text Table 16 show, the reidentification precision rate for nonmodal persons never exceeds 26.4\% and averages 20.1\% to 20.4\%. An attacker using the methods in this paper would be wrong four times in five guessing the \{\textit{race, ethnicity}\} from either the reconstructed MDF, $\text{rMDF}_{b,t}$, or the MDF itself. The PRG statistical baseline has reidentification precison of 16.0\% to 16.7\% on nonmodal persons (shown in supplemental text Table 13). The 2020 DAS therefore controls nonstatistical (potentially confidentiality-violating) inference at approximately the same rate as the proportional guesser.

Table \ref{tab:nm_solvar} \hypertarget{disc:nm_solvar}{contains the final piece of evidence} that the 2020 DAS, when applied to the 2010 Census data, effectively countered reconstruc\-tion-abetted reidentification attacks as we have modeled them here. For the nonmodal persons living in  zero solution variability blocks who are population unique on \{\textit{block, sex, agebin}\} (Panel B), the reidentification precision rates of $\text{rMDF}_{b,t}$ and MDF vary between 21.6\% and 21.8\%, essentially identical to the precision rates for the proportional guesser (20.9\% and 20.2\% for COMRCL and $\text{CEF}_{atkr}$, resp.). Neither the low nor high swap rate experiments deliver precision rates for this vulnerable population below 76.4\%. An attacker using $\text{rSWAPHi}_{b,t}$ and our methods would be correct about the population unique nonmodal \{\textit{race, ethnicity}\} approximately three times out of four, even though 50\% of the housing unit records have been swapped.

Because we have demonstrated, using 2010 Census data, that the tabular and microdata formats for publishing data processed by the 2020 DAS are properly protected either when reconstructed from DHC tables into $\text{rMDF}_{b,t}$ or as the MDF itself, both could be safely released. The 2020 DAS, therefore, demonstrably addresses the flawed assumption underlying the 2010 Census SDL framework---that tabular data products can be effectively protected using less stringent SDL methods than the microdata from which they are tabulated---by producing both tabular and microdata products within the same disclosure limitation framework using methods that ensure that the protections afforded to the tabular summaries and microdata are identical.

The 2020 DAS was designed to permit extensive tuning of the redistricting and DHC data products. Both were extensively tuned for accuracy, while still effectively defending against reconstruction-abetted re-identification attacks. The results of this accuracy tuning and corresponding measures of the data's anticipated fitness for use are available in other sources \citep{USCB:2023:DHC:DEMO:Metrics, Abowd20222020,cumings-menon:etal:2023} and in the replication archive for this paper. The replication archive also contains the same extensive set of metrics for $\text{rSWAPLo}_{b,t}$ and $\text{rSWAPHi}_{b,t}$. The low swap rate experiment has accuracy comparable to SF1. The high swap rate experiment has accuracy far worse than the production DHC on most statistics. 

The other feasible alternative to the 2020 DAS was a variant of the suppression methods used in the 1980 Census. Suppression can defend against an important subset of reconstruction attacks---isolating the records of population uniques; however, successful suppression systems, such as the one used for economic censuses \citep{cox:1995}, require both primary and complementary suppression. 

The supplemental text Appendix D lays out the rules used for the 1980 Census as we adapted them to testing suppression for the 2020 Census. Only primary and whole table suppression were used. Without complementary suppression there is no provable defense against reconstruction of population unique records. 

We considered only the four primary tables in the Redistricting Data (P.L. 94-171 ) Summary File (P8--P11 in Summary File 1). The supplemental text Tables 18 and 19 show that the use of just the 1980 Census primary suppression rules would have resulted in zeroing out up to 83.8\% of the cells in these tables,  suppressing up to 87.7\% of the block-level redistricting tables, and suppressing 38.7\% of the other block-level tables in the 2010 SF1. 

The use of suppression would, therefore, entirely preclude the redistricting use case---missing block-level tables prevent the creation of new voting districts, which have unknown boundaries at the time of data publication. The new districts must have approximately equal populations and be drawn in compliance with Section 2 of the 1965 Voting Rights Act. This limitation of suppression was known in 1980 and constituted one of the main reasons for using swapping in the 1990 Census data \citep{mckenna:2018:disclosure}.

Additional details and discussion can be found in the supplemental text Section 9.

\section{Conclusions}
\label{sec:conc}

This paper directly addresses the concept at the heart of traditional statistical disclosure limitation: the claim that as long as there is some uncertainty (``plausible deniability'' on the agency's part) regarding whether or not a particular respondent’s data were used in a particular statistic, then the agency can be considered to have provided meaningful confidentiality protection to respondents. 

We establish that the 2010 Census attacker does not need access to the confidential data to reconstruct a close approximation to the confidential HDF, to know that many records are identical in the reconstructed data and the HDF (using the 38-age-bin schema), and, in many cases, to know which records those are (see Section \ref{sec:solvar_results}). While this is strictly true for ``only'' 97 million persons, adding more tables to the reconstruction input data can only increase this number. If the reconstructed microdata meet the conditions for a microdata-based record-linkage attack, then such an attack must be defended in the SDL framework. Therefore, the plausible deniability argument turns on whether the 2010 Census aggregation and swapping were adequate. We argued here that they were not. In particular, if the released tabular data have reidentification precision rates that are essentially identical to the original confidential data, as shown in Table \ref{tab:nm_solvar}, then either those data could have been released without aggregation, which the Census Bureau has consistently argued would violate 13 U.S. Code §§ 8(b) \& 9 (Title 13, familiarly), or the tabular data are too disclosive, which is what we conclude. 

The results of our reidentification experiments demonstrate that the Census Bureau was correct in their assumption that the 2010 HDF required additional confidentiality protections beyond those applied to the tabular summaries in SF1. The record swapping mechanism used to produce the HDF is insufficient to prevent the disclosure of confidential census response information via reidentification attacks, especially for the most vulnerable populations. Moreover, by conducting these reidentification experiments on HDF microdata records reconstructed from published SF1 tables, these experiments further demonstrate the acute vulnerability of prior decades' SDL techniques, which treated tabular releases differently from microdata products, to reconstruction-abetted reidentification attacks. 

From the lens of why protections are necessary, a respondent who does not wish their or their child's race and ethnicity to be learned from the 2010 Census may understandably find our results concerning. They imply that a reconstruction-abetted attack could infer race and ethnicity with high confidence, and could only do so because the respondent provided this information to the Census Bureau. Although many other respondents may not find inference about race and ethnicity concerning, we remind the reader that race and ethnicity were used here as an example for illustration, not because of any special structure these variables have that is useful for a reconstruction-abetted reidentification attack. Presumably, many more respondents would be concerned if a similar attack could exploit their specific 2010 Census responses to infer whether they were in a same-sex marriage, whether their child was adopted, whether they were an older person living alone, whether they have exceeded their lease's occupancy limit, etc. 

This paper has also demonstrated that the decision to replace the SDL system used for the 2010 Census was based on sound scientific evidence that releasing large quantities of aggregate data, as the census has done for decades, is equivalent to releasing microdata. Further, we demonstrate that the methods used for the 2010 Census microdata would likely fail if applied to the 2020 Census tabular data and/or would lead to unacceptable degradation of data quality. 

First, we demonstrated that using only 5 billion of the 150 billion statistics published from the 2010 Census, the confidential microdata from the tabulation input file, the Hundred-percent Detail File, could be recreated with at least 95\% accuracy on the schema used for all block-level data. This image of the 2010 HDF, \rHDFBT, violates the more stringent release requirements for microdata products derived from the HDF, most notably that the records are not sampled and have geographic precision to the individual census block level rather than for geographies with a minimum population of 100,000. Similar criteria are still in place for public-use microdata samples for other Census Bureau household surveys. 

Second, we demonstrated that the reconstructed HDF microdata could be used to make high-precision confidentiality-violating inferences via reidentification attacks linking the reconstructed microdata to external attacker files. 

Finally, we demonstrated, by applying the 2020 DAS to 2010 Census data, that the 2020 DAS is able to successfully mitigate the threat of a reconstruction-abetted reidentification attack by greatly reducing the putative reidentification rates and by limiting the precision rates on vulnerable populations, while simultaneously producing statistics that meet established data quality requirements for their intended purposes. 

\subsection*{Disclosure Statement}
The research presented in this article was initiated, funded and supervised by the U.S. Census Bureau. All authors were either employees or contractors of the Census Bureau while performing their contributions.

\subsection*{Disclaimer and Acknowledgments}
The views and opinions expressed in this paper are those of the authors and not the U.S. Census Bureau. Research was conducted under Project ID: P-7502798. John Abowd, Tamara Adams, Robert Ashmead, Simson Garfinkel, Nathan Goldschlag, Philip Leclerc, Ethan Lew, Ramy Tadros, and Lars Vilhuber worked in their personal capacities and without access to any confidential data after leaving the Census Bureau (Abowd, Adams, Ashmead, Garfinkel, Goldschlag, Leclerc, and Vilhuber) and Galois (Lew and Tadros), respectively, to assist in preparing the manuscript for publication. Statistics reported were released under DRB Clearance numbers CBDRB-FY20-DSEP-001, CBDRB-FY22-DSEP-003, CBDRB-FY22-DSEP-004, and CBDRB‑FY23‑0152. The full technical report, referenced throughout this manuscript as the supplemental text, is available on ArXiv at  \url{https://doi.org/10.48550/arXiv.2312.11283}.

\subsection*{Data availability}
The public replication archive is located at \url{https://github.com/uscensusbureau/recon_replication}. The archive contains all tables and figures in this paper in an Excel\textsuperscript{TM} workbook. The archive also contains pointers to the official versions of all public data used in this paper, the complete code base, and a description of the workflow that meets current scientific best practices as defined by the American Economic Association \url{https://aeadataeditor.github.io/}. Census Bureau Data Stewardship Policy DS-027 \url{https://www2.census.gov/foia/ds_policies/ds027.pdf} permitted the editor and referees of journals that reviewed this work to independently verify the correctness of the calculations done on the confidential 2010 Census and commercial data used in this paper.

\subsection*{Contributions}
\begin{itemize}
\item Conceptualization: JMA, TA, SLG, DK, PL, LV
\item Data curation: NG, RAR, LV
\item Formal Analysis: TA, SLG, DK, PL
\item Methodology: JMA, TA, RA, DD, SD, SLG, NG, DK, PL, EL, SM, RAR, RNT, LV
\item Investigation: DD, SD, SLG, NG, DK, PL, EL, SM, RAR, RNT, LV
\item Software: TA, DD, SD, SLG, NG, PL, EL, SM, RAR, RNT, LV
\item Supervision: JMA
\item Writing – original draft: JMA, RA, SLG, NG, DK, PL, RAR, LV
\item Writing – review \& editing: JMA, TA, RA, DD, SD, SLG, NG, MBH, DK, PL, EL, SM, RAR, RNT, LV
\end{itemize}

\newpage


\printbibliography

\end{document}